\documentclass{lmcs}
\pdfoutput=1

\usepackage{silence}
\usepackage{lastpage}
\lmcsdoi{15}{3}{31}
\lmcsheading{}{\pageref{LastPage}}{}{}%
{Mar.~07,~2017}{Sep.~20,~2019}{}

\keywords{model-checking, bounded variable first-order logic, parameterized logarithmic space}

\newcommand{\npprob}[4]{
\begin{center}
\begin{tabular}{|r p{12.5cm} |} 
\hline 
\multicolumn{2}{|l|}{\emph{p}-\textsc{#1}}\\
\textit{Instance:}& #2.\\
\textit{Parameter:}& #3.\\
\textit{Problem:}& #4\\
\hline 
\end{tabular}
\end{center}
}

\newcommand{\nprob}[3]{\textup{
\begin{center}
\begin{tabular}{|r p{12.5cm} |} 
\hline 
\multicolumn{2}{|l|}{\textsc{#1}}\\
\textit{Instance:}& #2.\\
\textit{Problem:}& #3\\
\hline 
\end{tabular}
\end{center}
}}

\newcommand{\nnprob}[2]{\textup{
\begin{center}
\begin{tabular}{|r p{12.5cm} |} 
\hline 
\textit{Instance:}& #1.\\
\textit{Problem:}& #2\\
\hline 
\end{tabular}
\end{center}
}}

\newcommand{\A}{\mathbb{A}}
\newcommand{\B}{\mathbb{B}}
\newcommand{\N}{\mathbb{N}}

\newcommand{\pl}{\mathit{pl}}

\newcommand{\str}{\mathbf}
\newcommand{\cl}[1]{\textup{\sc #1}}

\newcommand{\FORALL}{\textbf{for all}}
\newcommand{\TO}{\textbf{to}}
\newcommand{\DO}{\textbf{do}}

\newcommand{\IF}{\textbf{if}}

\newcommand{\THEN}{\textbf{then}}

\newcommand{\im}[1]{\item\hspace{#1cm}}
\newenvironment{algorithm}{\begin{enumerate}
}{\end{enumerate}}

\newcommand{\paraL}{\textup{para-L}}
\newcommand{\paraNL}{\textup{para-NL}}

\newcommand{\parabetaL}{\textup{para}\beta\textup{L}}

\newcommand{\stdipath}{\textsc{stcon}}
\newcommand{\PATH}{\textup{PATH}}
\newcommand{\TREE}{\textup{TREE}}
\newcommand{\FPT}{\textup{FPT}}

\newcommand{\C}{\mathbb{C}}

\newcommand{\bigmid}{\;\big|\;}

\newcommand{\trans}{\textit{trans}}
\newcommand{\myvalue}{\textit{value}}
\newcommand{\tuple}{\textit{tuple}}

\newcommand{\benda}{\tag*{$\Box$}}

\newcommand{\FO}{\textup{FO}}

\newcommand{\STCon}{\textsc{stcon$_\le$}}

\newcommand{\pmc}{\textsc{$p$-mc}}

\usepackage{thm-restate}

\begin{document}

\title[Model-checking bounded variable first-order logic]{The parameterized space complexity of model-checking bounded variable first-order logic\rsuper*}
\titlecomment{{\lsuper*}Some of the results in this paper have appeared in the extended abstract~\cite{mfcs}.}

\author[Y. Chen]{Yijia Chen\rsuper{a,1}}
\address{\lsuper{a}School of Computer Science, Fudan University, China}
\email{yijiachen@fudan.edu.cn}
\thanks{\lsuper{1}Supported by the National Natural Science Foundation of China (NSFC) under project 61872092.}

\author[M. Elberfeld]{Michael Elberfeld\rsuper{b}}
\address{\lsuper{b}Lehrstuhl f\"ur Informatik 7, RWTH Aachen, Germany}
\email{elberfeld@informatik.rwth-aachen.de}

\author[M. M\"uller]{Moritz M\"{u}ller\rsuper{c,2}}
\address{\lsuper{c}Department of Computer Science, Universitat Polit\`ecnica de Catalunya, Spain}
\email{moritz@cs.upc.edu}
\thanks{\lsuper{2}Supported by the Austrian Science Fund (FWF) under project number P28699. The final revision has been done with support by the European Research Council (ERC) under the European Unions Horizon 2020 research programme (grant agreement ERC-2014-CoG 648276 AUTAR).}

\begin{abstract}
The parameterized model-checking problem for a class of first-order
sentences (queries) asks to decide whether a given sentence from the class
holds true in a given relational structure (database); the parameter is
the length of the sentence.
%
%
We study the parameterized space complexity of the model-checking problem
for queries with a bounded number of variables. For each bound on the
quantifier alternation rank the problem becomes complete for the
corresponding level of what we call the tree hierarchy, a hierarchy of
parameterized complexity classes defined via space bounded alternating
machines between parameterized logarithmic space and fixed-parameter
tractable time. We observe that a parameterized logarithmic space
model-checker for existential bounded variable queries would allow to
improve Savitch's classical simulation of nondeterministic logarithmic
space in deterministic space $O(\log^2n)$. Further, we define a highly
space efficient  model-checker for queries with a bounded number of
variables and bounded quantifier alternation rank. We study its optimality
under the assumption that Savitch's Theorem is optimal.
\end{abstract}

\maketitle

\section{Introduction}

The model-checking problem $\textsc{mc}(\text{FO})$ for first-order logic FO asks
whether a given first-order sentence $\varphi$ holds true in a given
relational structure $\str A$. This problem is of paramount importance
throughout computer science, and especially in database theory~\cite{sss}.
The problem is PSPACE-complete in general and even its restriction to
primitive positive sentences and two-element structures stays NP-hard
(cf.~\cite{ch}). Hence neither syntactic nor structural restrictions seem to
allow to get a handle on this problem. An important exception is the
observation that the natural bottom-up evaluation algorithm takes only
polynomial time on sentences with a bounded number of variables
(see~\cite[Proposition~3.1]{var95}, or the proof
of~\cite[Theorem~B.5]{immerman}). Indeed, this algorithm runs in time
$O(|\varphi|\cdot |\str A|^s)$ on instances $(\str A,\varphi)$ where
$\varphi$ is in $\cl{FO}^s$, i.\,e.\ contains at most~$s$ many variables.

Following Chandra and Merlin's seminal paper~\cite[Section~4]{ch}, it has
repeatedly been argued in the literature
(see~e.g.~\cite{sss,flumgrohemodcheck}) that measuring computational
resources needed to solve $\textsc{mc}(\text{FO})$ by functions in the length of the
input only is unsatisfactory. It neglects the fact that in typical situations
in database theory we are asked to evaluate a relatively short~$\varphi$ (the
query) against a relatively large $\str A$ (the database).
%
%
Parameterized complexity theory measures computational resources by
functions taking as an additional argument a \emph{parameter} associated to
the problem instance. For the parameterized model-checking problem
$\pmc(\FO)$ one takes the length $|\varphi|$ of~$\varphi$ as parameter and
asks for fixed-parameter tractable restrictions of $\pmc(\FO)$, i.\,e.\
restrictions decidable in \emph{fpt time} $f(|\varphi|)\cdot |\str A|^{O(1)}$
where $f:\N\to\N$ is an arbitrary computable function.

This relaxed tractability notion allows for effective but inefficient
formula manipulations and thereby the transfer of logical methods to the
study of the problem (see~\cite{grohemodcheck, grohesurvey, gksurvey} for
surveys). For example, a sequence of works constructing algorithms
exploiting Gaifman's locality theorem~\cite[Theorem~2.5.1]{fmt} recently led
to an fpt time algorithm solving $\pmc(\FO)$ on nowhere dense graph
classes~\cite{nowhere}.
For graphs of bounded degree one even gets an algorithm that runs in
parameterized logarithmic space~\cite{para}. Parameterized logarithmic
space, denoted by $\paraL$, relaxes logarithmic space in much the same way
as~FPT relaxes polynomial time~\cite{fellowscai, para}.

Concerning restrictions on the syntactical side one can naturally stratify
the problem into subproblems $\pmc(\Sigma_1)$, $\pmc(\Sigma_2), \ldots,
\pmc(\FO)$ according to quantifier alternation rank. These problems stay
intractable in the sense that they are well known to be complete for the
levels of the A-hierarchy $\cl{A}[1]\subseteq \cl{A}[2]\subseteq \cdots
\subseteq \cl{AW}[*]$. The completeness stays true even in the presence of
function symbols~\cite[p. 201]{flumgrohebuch}.

One of the main research questions is to understand for which sets~$\Phi$ of
first-order sentences, the problem $\pmc(\Phi)$ is tractable. In this
introductory exposition let us restrict attention to decidable sets $\Phi$
in a (relational) vocabulary that has bounded arity.

Today, the situation is well understood for sets $\Phi$ of primitive positive
sentences, i.\,e.\ \emph{conjunctive queries} in database
terminology~\cite{ch}. Indeed, under the assumption $\cl{A}[1]\neq\cl{FPT}$,
the model-checking problem $\pmc(\Phi)$ is fixed-parameter tractable
if~\cite{dalmau} and only if~\cite{gss,grohe} there is some constant $s\in\N$
such that every query in $\Phi$ is logically equivalent to a conjunctive
query whose primal graph has treewidth at most $s+1$. Now, these are
precisely those conjunctive queries that can be written with at most $s$ many
variables~\cite[Lemma~5.2, Remark~5.3]{kv00} (see
also~\cite[Theorem~12]{dalmau} and~\cite[Theorem~4]{hubie1} for similar
statements). In other words, if $\pmc(\Phi)$ is fixed-parameter tractable at
all, then it can be decided by first preprocessing the query to one with a
bounded number of variables and then run the natural evaluation algorithm. As
pointed out in~\cite{hubie1}, this is a recurring paradigm. In fact, most
known sets $\Phi$ (not necessarily primitive positive) with a tractable
$\pmc(\Phi)$ are contained in~$\cl{FO}^s$ up to logical equivalence.

This gives special interest to the computational complexity of
$\pmc(\cl{FO}^s)$. Moreover, already for $s=2$ this problem encompasses
problems of independent interest. For example, a directed graph with two
vertices named by constants $s$ and $t$ contains a path from $s$ to $t$ of
length at most~$k$ if and only if it satisfies the sentence $\varphi_k(s)\in
\cl{FO}^2$ (allowing constants~$s,t$) where $\varphi_k(x)$ is defined as
follows:
\begin{equation}\label{eq:fostcon}
\begin{array}{rl}
\varphi_0(x) &:= x{=}t, \\[1mm]
\varphi_{k+1}(x)
 &:= \exists y \Big(\big(y{=}x\vee Exy\big) \wedge \exists x\big(x{=}y \wedge \varphi_{k}(x)\big)\Big)
\end{array}
\end{equation}

%
%

There is some recent work concerning the fine-grained time complexity of
$\pmc(\cl{FO}^s)$, see~\cite{fine}. But given the central importance of
$\pmc(\cl{FO}^s)$ it seems surprising that, to the best of our knowledge, its
space complexity has not been thoroughly studied. It is known that
$\textsc{mc}(\cl{FO}^s)$ is P-complete for $s\ge 2$ under logarithmic space
reductions (see~\cite{immerman,var95}).
But, even
assuming~$\cl{P}\neq\cl{L}$, this leaves open the possibility that
$p$-\textsc{mc}$(\cl{FO}^s)$ could be solved in \emph{parameterized
logarithmic space}, that is, (deterministic) space $f(|\varphi|)+O(\log|\str
A|)$ for some computable $f:\N\to\N$. The central question of this paper is
whether this is the case.\footnote{Relatedly, it has been asked by Flum and
Grohe~\cite[Remark~26]{para} whether $\pmc(\cl{FO}^s)$ is in $\paraNL$.}

As we shall see, answering this question either positively or negatively
would have breakthrough consequences in classical complexity theory. It is
one of the central open questions or, in Lipton's
words~\cite[p.137]{liptonbook}, ``one of the biggest embarrassments of all
complexity theory'' whether Savitch's upper bound $\cl{NL}\subseteq
\cl{SPACE}(\log^2n)$ can be improved. This is open since 1969 and there is a
tangible possibility that \emph{Savitch's Theorem is optimal},
i.\,e.\footnote{By the space hierarchy theorem $\cl{SPACE}(s(n))\neq
\cl{SPACE}(\log^2 n)$ holds for the Turing machine model and for all (not
necessarily space constructible) $s(n)\le o(\log^2 n)$.}
\begin{equation*}
\cl{NL}\not\subseteq\cl{SPACE}(o(\log^2n)).
\end{equation*}
See~\cite{hem,pot10} for more information about this problem.
%
We observe the following implications:
\begin{equation*}\label{eq:impl}
\cl{P}=\cl{L}
  \quad \Longrightarrow\quad
 \pmc(\cl{FO}^s)\in\paraL
  \quad\Longrightarrow\quad
 \textup{Savitch's Theorem is not optimal}.
\end{equation*}
For the second implication, note that running the assumed model-checker on
the sentences~\eqref{eq:fostcon} solves the parameterized problem
\npprob{stcon$_\le$}{A directed graph $\str G$, two vertices $s$ and $t$,
and $k\in \N$}{$k$}{Is there a directed path of length at most $k$ from $s$
to $t$ in $\str G$?}
On $n$-node graphs this requires only space $f(k)+O(\log n)$ for some
computable $f$. Then there is a logspace computable unbounded function $h$
such that the algorithm runs in logarithmic space on instances with $k\le
h(n)$. It is well known that such an algorithm implies that Savitch's theorem
is not optimal (see e.g.~\cite{wigderson}). We shall give the details of
this argument, and in fact prove something stronger. More precisely, the
contributions of this paper are as follows.

\paragraph{\bf Our contributions.}
Our central question asks for the complexity of $\pmc(\cl{FO}^s)$ up to
parameterized logarithmic space reductions, i.\,e.\ \emph{pl-reductions}. An
obvious approach is to stratify the problem again according to quantifier
alternation rank into subproblems $\pmc(\Sigma_1^s)$,
$\pmc(\Sigma_2^s),\ldots$. It turns out that there exist pl-reductions from
$\pmc(\cl{FO}^s)$ to~$\pmc(\cl{FO}^2)$,
and for each fixed $t$ from $\pmc(\Sigma_t^s)$ to $\pmc(\Sigma_t^2)$ 
(see Theorem~\ref{thm:completeness} below). On the other hand we do not know
whether one can reduce $\pmc(\Sigma_{t+1}^2)$ to $\pmc(\Sigma_t^s)$ for any
$s$.

Clearly, all these problems lie between $\paraL$ and \FPT\@. But unfortunately
not much is known about this complexity landscape, and only recently there
is a slowly emerging picture~\cite{est,cm1,cm2,mfcs}. In particular,~\cite{est} introduced a parameterized analogue of NL, called \PATH\
in~\cite{cm1}, and~\cite{cm1} introduced the class \TREE, a parameterized
analogue of LOGCFL\@. Unlike their classical counterparts~\cite{sze,imm,cook},
$\PATH$ and $\TREE$ are not known to be closed under complementation (this
is a question from~\cite{cm1}). We thus face the naturally defined
alternation hierarchy built above~\TREE\ (see~Definition~\ref{df:tree}). We
call it the \emph{tree hierarchy}:
\begin{equation}\label{eq:hierarchy}
\paraL\subseteq \PATH \subseteq \TREE= \TREE[1] \subseteq\TREE[2]
 \subseteq \cdots \subseteq
\TREE[*]\subseteq \FPT.
\end{equation}

Our first result reads as follows:
\begin{thm}\label{thm:completeness} Let $s\ge 2$ and $t\ge 1$.
\begin{enumerate}
\item  $\pmc(\Sigma^s_t)$ is complete for $\cl{Tree}[t]$ under
    pl-reductions.

\item $\pmc(\cl{FO}^s)$ is complete for $\cl{Tree}[*]$ under
    pl-reductions.
\end{enumerate}
\end{thm}

\noindent
We shall also prove that these results stay true in the presence of function
symbols (Theorem~\ref{thm:func}). This situation within \FPT\ is thus fully
analogous to the situation with unboundedly many variables and the
A-hierarchy. The proofs, however, are quite different. A further difference
is that the tree hierarchy satisfies a collapse theorem
(Corollary~\ref{cor:collapse}) like the polynomial hierarchy while such a
theorem is unknown for the A-hierarchy.

The connection to the classical question whether Savitch's Theorem is
optimal reads, more precisely, as follows:

\begin{restatable}{thm}{improvesavitch}\label{thm:improvesavitch}
If Savitch's  Theorem is optimal, then  $ \PATH\not\subseteq\paraL$.
\end{restatable}

Concerning our central question, these results together
with~\eqref{eq:hierarchy} imply that, if Savitch's Theorem is optimal, then
already the lowest level $\pmc(\Sigma_1^2)$ cannot be solved in
parameterized loga\-rithmic space. In fact, we show something stronger:

\begin{restatable}{thm}{thmlower}\label{thm:lower}
Let $f$ be an arbitrary function from $\N$ to $\N$. If Savitch's Theorem is
optimal, then $\pmc(\Sigma_1^2)$ is not decidable in space
$o(f(|\varphi|)\cdot \log |\str A|)$.
\end{restatable}

A straightforward algorithm solves $\pmc(\cl{FO}^s)$ in space
(cf.~Lemma~\ref{lem:oed})
\[
O(|\varphi|\cdot (\log|\varphi|+\log|\str A|)).
\]
Theorem~\ref{thm:lower} shows that there might be not too much room for
improvement. However, building on ideas of Ruzzo~\cite{ruzzo}, we still
manage to give a significant improvement for bounded quantifier alternation
rank:

\begin{restatable}{thm}{thmupper}\label{thm:upper}
For all $s,t\in\N$ the problem $\pmc(\Sigma_t^s)$ is decidable in space
\[
O(\log|\varphi|\cdot (\log|\varphi|+\log|\str A|)).
\]
\end{restatable}

It is unlikely that the bound on the quantifier alternation rank can be
omitted in this statement. Indeed, it follows from
the P-completeness
result mentioned above that $\pmc(\cl{FO}^2)$ cannot be decided in the
displayed space unless $\cl{P}\subseteq \cl{SPACE}(\log^2 n)$.

\section{Preliminaries}

For a natural number $n\in \N$ we set $[n]= \{1, \ldots, n\}$ understanding
$[0]= \emptyset$.

\subsection{Structures}
A \emph{vocabulary} $\tau$ is a finite set of relation and function symbols.
Relation and function symbols have an associated \emph{arity}, a natural
number. A \emph{$\tau$-structure $\str A$} consists of a finite nonempty set
$A$, its \emph{universe}, and for each $r$-ary relation symbol $R\in\tau$ an
\emph{interpretation} $R^{\str A}\subseteq A^r$ and for each $r$-ary function
symbol $f\in \tau$ an \emph{interpretation} $f^{\str A}: A^r\to A$. A \emph{constant} is a function symbol $c$ of arity 0. We identify its
interpretation $c^{\str A}$ with its unique value, an element of $A$.

A \emph{directed graph} is an $\{E\}$-structure $\str G=(G,E^{\str G})$ for
the binary relation symbol $E$. We refer to elements of $G$ as \emph{vertices} and to elements  $(a,b)\in E^{\str G}$ as \emph{(directed) edges
(from~$a$ to~$b$)}. Note this allows $\str G$ to have self-loops, i.\,e.\
edges from $a$ to $a$. A \emph{graph} is a directed graph $\str G=(G,E^{\str
G})$ with irreflexive and symmetric $E^{\str G}$. A \emph{(directed) path} in
a (directed) graph $\str G$ is a sequence $(a_1,\ldots, a_{k+1})$ of
pairwise distinct vertices such that $(a_i,a_{i+1})\in E^{\str G}$ for all
$i\in[k]$; the path is said to have \emph{length $k$} and to be \emph{from
$a_1$ to $a_k$}. Note that there is a path of length 0 from every vertex to
itself.

The \emph{size} of a $\tau$-structure $\str A$ is
\[
\textstyle
|\str A|:= |\tau|+ |A|+ \sum_{R} |R^{\str A}|\cdot
\textit{ar}(R)+ \sum_{f} |A|^{\textit{ar}(f)},
\]
where $R,f$ range over the relation and function symbols of $\tau$
respectively, and $\textit{ar}(R)$, $\textit{ar}(f)$ denote the arities of
$R$, $f$ respectively. For example, the size of a (directed) graph with~$n$
vertices and $m$ edges is $O(n+m)$. Note that a reasonable binary
(``sparse'' or ``list'') encoding of $\str A$ has length $O(|\str A|\cdot
\log|A|)$. The difference between the size as defined and the length of the
binary encoding of a structure plays no role in this paper.

\subsection{Formulas}
Let $\tau$ be a vocabulary. A \emph{$\tau$-term} is a variable, or of the
form $ft_1\cdots t_{r}$ where $f$ is an $r$-ary function symbol and $t_1,
\ldots, t_r$ are again $\tau$-terms. \emph{Atomic} $\tau$-formulas, i.\,e.\
\emph{$\tau$-atoms},  have the form $t{=} t'$ or $R(t_1, \ldots, t_r)$ where
$R$ is an $r$-ary relation symbol in $\tau$ and $t, t', t_1, \ldots, t_r$
are $\tau$-terms. General \emph{$\tau$-formulas} are built from atomic ones
by $\wedge,\vee,\neg$ and universal and existential quantification $\forall
x,\exists x$. The vocabulary $\tau$ is \emph{relational} if it contains only
relation symbols. For a tuple of variables $\bar x=(x_1,\ldots, x_k)$ we
write $\varphi=\varphi(\bar x)$ to indicate that the free variables of
$\varphi$ are among $\{x_1,\ldots,x_k\}$. If $\str A$ is a $\tau$-structure
and $\bar a=(a_1,\ldots, a_k)\in A^k$, then $\str A\models\varphi(\bar a)$
means that the assignment that maps $x_i$ to $a_i$ for~$i\in[k]$ satisfies
$\varphi$ in $\str A$. A \emph{sentence} is a formula without free variables.
The \emph{size} $|\varphi|$ of a formula $\varphi$ is the length of a
reasonable binary encoding of it.

For $s\in \N$ let $\cl{FO}^s$ denote the class of (first-order) formulas
\emph{over a relational vocabulary} containing at most $s$ variables (free or
bound). For $t\in \N$ we define the classes $\Sigma_t$ and $\Pi_t$ as
follows. Both $\Sigma_0$ and $\Pi_0$ are the class of quantifier free
formulas; $\Sigma_{t+1}$ (resp.\ $\Pi_{t+1}$) is the closure of $\Pi_t$
(resp.\ $\Sigma_t$) under positive Boolean combinations (i.\,e.\ applying
$\vee,\wedge$) and existential (resp.\ universal) quantification. We set
\begin{enumerate}
\item[--] $\Sigma_t^s:= \cl{FO}^s\cap \Sigma_t$,

\item[--] $\Pi_t^s:= \cl{FO}^s\cap \Pi_t$.
\end{enumerate}

\begin{exa}\label{ex:fostconrel}
The formulas~\eqref{eq:fostcon} in the introduction do not qualify as
$\FO^2$ because they use constants $s,t$. They are evaluated in structures
$\str G=(G,E^{\str G},s^{\str G},t^{\str G})$ where $(G,E^{\str G})$ is a
directed graph and $s^{\str G},t^{\str G}\in G$ are two vertices. Let $S,T$
be unary relation symbols and consider the $\{E,S,T\}$-structure $\str
G'=(G,E^{\str G'},S^{\str G'},T^{\str G'})$ with $E^{\str G'}:=E^{\str
G},S^{\str G'}:=\{s^{\str G}\}$ and $T^{\str G'}:=\{t^{\str G}\}$. Define
formulas $\varphi_k'(x)$ as $\varphi_k(x)$ but with $\varphi'_0(x):=T(x)$.
Then $\exists x(S(x)\wedge\varphi'_k(x))$ is in $\Sigma^2_1$ and
\[
\str G\models\varphi_k(s) \iff \str G'\models\exists x(S(x)\wedge\varphi'_k(x)).
\]
\end{exa}

\subsection{Parameterized complexity}\label{subsec:complexity}
A \emph{(classical) problem} is a subset $Q\subseteq {\{0,1\}}^*$, where ${\{0,1\}}^*$ is the set of finite binary strings; the length of a binary
string $x$ is denoted by $|x|$. As model of computation we use Turing
machines~$\A$ with a (read-only) input tape and several worktapes. We shall
consider Turing machines with nondeterminism and co-nondeterminism. For
definiteness, let us agree that a \emph{nondeterministic} Turing machine has
special states~$c_\exists, c_0, c_1$ and can nondeterministically move from
state $c_\exists$ to state $c_b$ with $b\in \{0,1\}$, and we say $\A$
\emph{existentially guesses the bit $b$}. An \emph{alternating} Turing
machine additionally has a state $c_\forall$ allowing to \emph{universally
guess} a bit $b$. For a function $c: {\{0,1\}}^*\to \N$, the machine is said
to \emph{use $c$ many (co-)nondeterministic bits} if for every $x\in {\{0, 1\}}^*$ every run\footnote{By a \emph{run} of an alternating Turing machine we 
mean a sequence of configurations such that each, except the first, is a
successor configuration of the previous one.} of $\A$ on $x$ contains at
most $c(x)$ many configurations with state $c_\exists$ (resp. $c_\forall$).

A \emph{parameterized problem} is a pair $(Q,\kappa)$ of a classical problem
$Q$ and a \emph{parameterization}~$\kappa$, i.\,e.\ a logarithmic space
computable function $\kappa: {\{0,1\}}^* \to \N$ mapping any  $x\in
{\{0,1\}}^*$ to its \emph{parameter} $\kappa(x) \in \N$.

We exemplify how we present parameterized problems. The \emph{model-checking
problem for} a class of first-order sentences $\Phi$ is the parameterized
problem
\npprob{mc$(\Phi)$}{A first-order sentence $\varphi$ and a structure $\str
A$}{$|\varphi|$} {$\varphi\in\Phi$ and $\str A\models \varphi$?}
More formally, this is the classical problem $\textsc{mc}(\Phi)$ containing all
(binary strings encoding) pairs $(\varphi, \str A)$ with $\varphi\in\Phi$ and
$\str A\models \varphi$, together with a parameterization that maps binary
strings encoding pairs of formulas and structures to the length of the binary
string encoding the formula, and all other strings to 0.

The class \FPT\ contains those parameterized problems $(Q,\kappa)$ that can
be decided in \emph{fpt time with respect to $\kappa$}, i.\,e.\ in time
$f(\kappa(x))\cdot |x|^{O(1)}$ for some computable  $f: \N\to \N$. The class
$\paraL$ ($\paraNL$) contains those parameterized problems $(Q,\kappa)$ such
that $Q$ is decided (accepted) by some (nondeterministic) Turing machine
$\A$ that runs in \emph{parameterized logarithmic space with respect to
$\kappa$}, i.\,e.\ in space $f(\kappa(x))+ O(\log |x|)$ for some computable
function $f: \N\to \N$. We remark that the class~XL is defined using space bound
$f(\kappa(x))\cdot\log|x|$ instead $f(\kappa(x))+ O(\log |x|)$. This class is not known to be contained in \FPT\@.
We shall omit the phrase ``with respect to $\kappa$'' if $\kappa$ is clear
from context.

\emph{Parameterized logarithmic space reductions} have been introduced in
~\cite{para}. We use the following equivalent definition: a
\emph{pl-reduction} from $(Q,\kappa)$ to $(Q',\kappa')$ is a reduction
$R:\{{0,1\}}^*\to {\{0,1\}}^*$ from $Q$ to $Q'$ such that $\kappa'(R(x))\le
f(\kappa(x))$ and $|R(x)|\le f(\kappa(x))\cdot |x|^{O(1)}$ for some
computable function $f: \N\to \N$, and $R$ is \emph{implicitly
pl-computable}, that is, the following problem is in \paraL:\@
\npprob{Bitgraph$(R)$}{$(x, i, b)$ with $x\in {\{0,1\}}^*$, $i\ge 1$, and
$b\in\{0,1\}$}{$\kappa(x)$}{Does $R(x)$ have length $|R(x)| \geq i$ and
$i$-th bit $b$?}
If there is such a reduction, $(Q,\kappa)$ is \emph{pl-reducible} to
$(Q',\kappa')$, written $(Q,\kappa)\le_\pl (Q',\kappa')$.  If also
$(Q',\kappa)\le_\pl (Q,\kappa)$, then the problems are \emph{pl-equivalent},
written $(Q,\kappa)\equiv_\pl (Q',\kappa')$.

It is routine to verify that $\le_\pl$ is transitive and $\equiv_\pl$ an
equivalence relation.

\subsection{The classes \texorpdfstring{$\PATH$}{PATH} and \texorpdfstring{$\TREE$}{TREE}}\label{sec:pathandtree}

The class $\PATH$ has been introduced in~\cite{est} (where it is called
$\parabetaL$) and can be viewed as a parameterized analogue of~NL\@:

\begin{defi}
 A parameterized problem $(Q,\kappa)$ is in $\PATH$ if and only if there
exists a nondeterministic Turing machine that accepts $Q$, runs in
parameterized logarithmic space, and uses $f(\kappa(x))\cdot \log|x|$ many
nondeterministic bits for some computable $f: \N\to \N$.
\end{defi}

\begin{exa}
To gain some intuition for this definition consider the \emph{homomorphism
problem for directed paths}:\footnote{The homomorphism problems for directed
paths and binary trees (see Example~\ref{exa:treehom}) are known to be
$\PATH$- and $\TREE$-complete under pl-reductions, respectively~\cite{cm1}.}
the input is a directed path $\str P$ and a directed graph $\str G$; the
question is whether there is a homomorphism from $\str P$ into $\str G$; the
parameter is $k:=|P|$.

Note that $k\cdot \log|G|$ nondeterministic bits are enough to guess a
solution, they can however not be stored in parameterized logarithmic space.
However, the guess and check algorithm can nevertheless be implemented in
such space by observing that, intuitively, a solution can be verified
locally. More precisely,  the algorithm guesses ($\log |G|$ bits to
determine) $b\in G$ and writes~$(a,b)$ on some tape where $a$ is the source
of $\str P$. It then repeatedly updates this tape as follows:  accept if $a$
is the sink of $\str P$; else guess $b'\in G$ and  check $(b,b')\in E^{\str
B}$; if this fails, reject; otherwise replace $(a,b)$ by $(a',b')$ where $a'$
is the successor of $a$ in $\str P$.
\end{exa}

One similarly verifies that $p$-$\STCon$ is in $\PATH$. In fact, as has been
shown in~\cite[Theorem~3.14]{est}:

\begin{thm}\label{thm:pathcomplete}
The parameterized problem $p$-$\STCon$ is $\PATH$-complete under
pl-reductions.
\end{thm}

The class $\TREE$ has been introduced in~\cite{cm1} and can be viewed as a
parameterized analogue of LOGCFL:\footnote{The analogy is based on the
characterization of LOGCFL as logspace uniform SAC$_1$~\cite{venka}.}

\begin{defi}
A parameterized problem $(Q,\kappa)$ is in $\TREE$ if and only if there
exists an alternating Turing machine that accepts $Q$, runs in parameterized
logarithmic space, and for some computable $f: \N\to \N$ uses
$f(\kappa(x))\cdot \log|x|$ many nondeterministic bits and $f(\kappa(x))$
many co-nondeterministic bits.
\end{defi}

\begin{exa}\label{exa:treehom}
Again to gain some intuition consider the \emph{homomorphism problem for
directed binary trees}: the input is a full binary tree $\str T$ with edges
directed away from the root, and a directed graph $\str G$; the question is
whether there is a homomorphism from $\str T$ into $\str G$; the parameter is
$k:=|T|$.

To see that this problem belongs to $\TREE$ we note that a solution can be
locally verified with the help of universal guesses, namely with $h\cdot \log
k$ many co-nondeterministic bits where $h:=\log k -1$ is the height of $\str
T$. As in the previous example the algorithm maintains a pair $(a,b)\in
T\times G$ on some tape, starting with $a$ being the root of $\str T$. The
tape is updated as follows: if~$a$ is a leaf of $\str T$, accept; otherwise
universally guess ($\log k$ bits to deter\-mine)~$a'\in T$ and check
$(a,a')\in E^{\str T}$; if this fails, accept; else existentially guess
$b'\in G$ and check $(b,b')\in E^{\str G}$; if this fails, reject; else
replace $(a,b)$ by $(a',b')$.
\end{exa}

As outlined in the introduction the classes \PATH\ and \TREE\ play a central
role in this article. For this reason we provide more background on these
classes in the following by explaining their key role to understanding the
parameterized space complexity of the model-checking problem for structurally
restricted primitive positive sentences. We restate the relevant results
reformulated in our context. They are not needed later on.


Recall that a sentence $\varphi$ in a relational vocabulary is \emph{primitive positive} if it is built from atoms by means of conjunctions and
existential quantifications. To define the \emph{primal graph}~$\str
G(\varphi)$ first write $\varphi$ in prenex form (introducing new variables)
and then delete each subformula of the form $x{=}y$ replacing all
occurrences of $y$ by $x$; the graph $\str G(\varphi)$ has  the variables of
the resulting sentence as vertices, and an edge between two distinct
variables if they occur both in some atomic subformula of $\varphi$. A class
$\Phi$ of primitive positive sentences has \emph{bounded arity} if there is a
constant $r\in\N$ such that all relation symbols in all sentences in $\Phi$
have arity at most $r$. Let us say, $\Phi$ has \emph{bounded tree-depth up to
logical equivalence} if there is a constant $c\in\N$ such that every
sentence in $\Phi$ is logically equivalent to a primitive positive sentence
whose primal graph has tree-depth at most~$c$. We use a similar mode of
speech for path-width and tree-width.

Then the complexity classification of problems $\pmc(\Phi)$ reads as
follows.

\begin{thm}\label{theo:class}
Let $\Phi$ be a decidable class of primitive positive sentences of bounded
arity.
\begin{enumerate}

\item If $\Phi$ has bounded tree-depth up to logical equivalence, then
    $\pmc(\Phi)\in \paraL$.

\item If $\Phi$ does not have bounded tree-depth but bounded path-width up
    to logical equivalence, then $\pmc(\Phi)$ is $\PATH$-complete under
    pl-reductions.

\item If $\Phi$ does not have bounded path-width but bounded tree-width up
    to logical equivalence, then $\pmc(\Phi)$ is $\TREE$-complete under
    pl-reductions.

\item If $\Phi$ does not have bounded tree-width up to logical
    equivalence, then $\pmc(\Phi)$ is not in $\textup{FPT}$ unless
    $\textup{A[1]}= \textup{FPT}$.

\end{enumerate}
\end{thm}

\noindent
Statement (4) is Grohe's famous classification~\cite{grohe} (building on~\cite{gss}). The other statements are from~\cite{cm1}. An even finer
classification (including (4)) appears in~\cite{cm2}. Note statements (3)
and~(4) state a complexity gap (assuming $\textup{A[1]}\neq \textup{FPT}$):
if $\pmc(\Phi)$ is in \FPT\ at all, then it is already in $\TREE$.

Originally these results have been phrased for homomorphism problems. The
equivalence of the problems is well known since Chandra
and Merlin's seminal paper~\cite{ch}. Instead of logical equivalence of
sentences one talks about homomorphic equivalence of relational structures,
that is, structures with isomorphic cores.

Finally, the connection to bounded variable logics mentioned in the
introduction follows from~\cite[Lemma~5.2, Remark~5.3]{kv00} (see also~\cite[Theorem~12]{dalmau} and~\cite[Theorem~4]{hubie1} for similar
statements):

\begin{thm}
Let $s\in\N$ and $\varphi$ be a primitive positive sentence. Then $\varphi$
is logically equivalent to a primitive positive sentence whose primal graph
has treewidth less than $s$ if and only if $\varphi$ is logically equivalent
to a primitive positive sentence in $\FO^s$.
\end{thm}

\section{The tree hierarchy}\label{sec:nondet}

Theorem~\ref{theo:class} and the following discussion give some special
interest to the class \TREE\@. It has been asked in~\cite{cm1} whether $\TREE$
is closed under complementation. If not, then we get an alternation
hierarchy above \TREE\ defined in the usual way (that still might collapse
to a level further up). In this section we define this \emph{tree hierarchy}
and make some initial observations.

\subsection{Definitions}

Following~\cite{cm1} we consider machines $\A$ with \emph{mixed
nondeterminism}. Additionally to the binary nondeterminism embodied in the
states $c_\exists,c_\forall,c_0,c_1$ from Section~\ref{subsec:complexity}
they use \emph{jumps} explained as follows. Recall that our Turing machines
have an input tape. During a computation on an input~$x$ of length $n:=|x|>0$
the cells numbered 1 to $n$ of the input tape contain the $n$ bits of $x$.
The machine has an \emph{existential} and a \emph{universal jump state}
$j_{\exists}$ resp. $j_{\forall}$. A successor configuration of a
configuration in a jump state is obtained by changing the state to the
initial state and placing the input head on an arbitrary cell holding an
input bit; the machine is said to \emph{existentially resp.\ universally jump
to} the cell.

Acceptance is defined as usual for alternating machines. Call a
configuration \emph{universal} if it has state $j_{\forall}$
or~$c_{\forall}$, and otherwise \emph{existential}. The machine $\A$ accepts
$x\in{\{0,1\}}^{*}$ if its initial configuration on $x$ is \emph{accepting}.
The set of accepting configurations\label{page:alt} is the smallest set that
contains all accepting halting configurations, that contains an existential
configuration if it contains at least one of its successor configurations,
and that contains a universal configuration if it contains all of its
successor configurations.

\medskip
Observe that the number of the cell to which the machine jumps can be
computed in logarithmic space by moving the input head stepwise to the left.
Intuitively, a jump should be thought as a guess of a number in $[n]$.


\medskip
Each run of $\A$ on some input $x$ contains a subsequence of jump
configurations (i.\,e.\ with state~$j_\exists$ or~$j_\forall$). For a
natural number $t\ge 1$ the run is \emph{$t$-alternating} if this
subsequence consists of~$t$ blocks, the first consisting of existential
configurations, the second in universal configurations, and so on. The
machine $\A$ is \emph{$t$-alternating} if every run of $\A$ on any input is
$t$-alternating.

Note that a 1-alternating machine can existentially jump but not universally;
it may however use universal (and existential) bits; in fact, the use of
nondeterministic bits is completely neglected by the above definition.

For $f: {\{0,1\}}^*\to \N$, we say $\A$ \emph{uses $f$ jumps (bits)} if
for every $x\in{\{0,1\}}^*$ every run of~$\A$ on~$x$ contains at most
$f(x)$ many jump configurations (resp.\ configurations with state~$c_\exists$
or~$c_\forall$).
As for a more general notation, note that every run of~$\A$ on~$x$ contains
a (possibly empty) sequence of \emph{nondeterministic configurations}, i.\,e.\
with state in $\{j_\exists,j_\forall,c_\exists,c_\forall\}$. The
\emph{nondeterminism type} of the run is the corresponding word over the
alphabet $\{j_\exists, j_\forall, c_\exists, c_\forall\}$.
For example, being $2t$-alternating means
having nondeterminism type in
${({\{j_\exists,c_\exists, c_\forall\}}^* {\{j_\forall, c_\exists,
c_\forall\}}^*)}^{t}$. Here and below, we use regular expressions to denote
languages over $\{j_\exists, j_\forall, c_\exists, c_\forall\}$.

\begin{defi}\label{df:tree}
 A parameterized problem $(Q,\kappa)$ is in $\cl{Tree}[*]$ if there are a
computable $f: \N\to \N$ and a machine $\A$ with mixed nondeterminism that
accepts~$Q$, runs in parameterized logarithmic space (with respect to
$\kappa$) and uses $f\circ \kappa$ jumps and $f\circ \kappa$ bits. If
additionally $\A$ is $t$-alternating for some $t\ge 1$, then $(Q, \kappa)$ is
in $\cl{Tree}[t]$.
\end{defi}

The definition of $\TREE[t]$ is due to Hubie Chen (personal communication).

\subsection{Observations}

The following two propositions are straightforward
(cf.~\cite[Lemmas~4.5,~5.4]{cm1}):

\begin{prop}\label{prop:pathjump}
A parameterized problem $(Q,\kappa)$ is in $\PATH$ if and only if there are
a computable $f:\N \to \N$ and a 1-alternating machine with mixed
nondeterminism that accepts~$Q$, runs in parameterized logarithmic space
(with respect to $\kappa$) and uses $f\circ \kappa$ jumps and $0$ bits.
\end{prop}

\begin{prop}\label{prop:tree1}
$\TREE= \TREE[1]$.
\end{prop}

Hence, all inclusions displayed in~\eqref{eq:hierarchy} in the introduction
are trivial except possibly the last one $\TREE[*]\subseteq \FPT$. We prove
it in Corollary~\ref{cor:treefpt}. It is likely to be strict:

\begin{prop}\label{prop:nlvstreestar} \
\begin{enumerate}
\item $\paraNL\subseteq \TREE[*]$ if and only if $\cl{NL}\subseteq
    \cl{L}$.

\item $\cl{FPT}\subseteq \TREE[*]$ if and only if $\cl{P}\subseteq
    \cl{L}$.
\end{enumerate}
\end{prop}

\begin{proof}
We prove (1), the proof of (2) is similar. For the backward direction, note that
$\cl{NL}\subseteq \cl{L}$ implies $\paraNL\subseteq \paraL$ by general
results of Flum and Grohe~\cite[Theorem~4, Proposition~8]{para}. The same
results also imply the forward direction noting that $\TREE[*]$ is contained
in XL\@.
We include a direct argument: assume $\paraNL\subseteq \TREE[*]$, let $Q$ be
a classical problem which is $\cl{NL}$-complete under logarithmic space
reductions and let $\kappa_0$ be the parameterization which is constantly
$0$. Then $(Q,\kappa_0)\in \paraNL\subseteq \TREE[*]$, so there are a
function $f$ and machine $\A$ with mixed nondeterminism that accepts $Q$ and
on input $x$ uses $f(0)+O(\log|x|)$ space and $f(0)$ jumps and bits. Thus,
$Q$ can be decided in logarithmic space by simulating $\A$ for all possible
outcomes of these constantly many jumps and bits.
\end{proof}

\begin{rem}~\cite{est} observed that $\paraNL\subseteq \PATH$ is equivalent to
$\cl{NL}\subseteq \cl{L}$.
\end{rem}

The following technical lemma will prove useful in the next section.

\begin{lem}[Normalization]\label{lem:normal}
Let $t\ge 1$ and $(Q,\kappa)$ be a parameterized problem.
\begin{enumerate}
\item $(Q,\kappa)\in \cl{Tree}[t]$ if and only if there are a computable
    $f:\N\to\N$ and a $t$-alternating machine $\A$ with mixed
    nondeterminism that accepts~$Q$, runs in parameterized logarithmic
    space and such that for all $x\in{\{0,1\}}^*$
    every run of $\A$ on $x$ has nondeterminism type:
    \begin{equation}\label{eq:type}
    {\big(\ {(j_\exists c_\forall)}^{f(\kappa(x))}
     {(j_\forall c_\exists)}^{f(\kappa(x))}\ \big)}^{\lfloor t/2\rfloor}
    {(j_\exists c_\forall)}^{{f(\kappa(x))}\cdot (t\ \mathrm{mod}\ 2)}. 
    \end{equation}

\item $(Q,\kappa)\in \cl{Tree}[*]$ if and only if there are a computable
    $f: \N\to \N$ and machine $\A$ with mixed nondeterminism that
    accepts~$Q$, runs in parameterized logarithmic space and such that for
    all $x\in{\{0,1\}}^*$ every run of $\A$ on $x$ has nondeterminism type:
    \begin{equation}\label{eq:typestar}
    {(j_\exists j_\forall)}^{f(\kappa(x))}.
    \end{equation}
\end{enumerate}
\end{lem}


\noindent \textit{Proof.} We only show (1), the proof of (2) is similar. The
backward direction of (1) is obvious. To prove the forward direction, assume
$(Q,\kappa)\in \cl{Tree}[t]$ and choose a parameterized logarithmic space
$t$-alternating machine $\A$ with mixed nondeterminism accepting $Q$ and a
computable function $f:\N\to \N$ such that $\A$ on input $x\in{\{0,1\}}^*$
uses $f(\kappa(x))$ many jumps and bits.

Every run of $\A$ on $x$ consists of at most $t$ \emph{blocks}. The first
block is the sequence of configurations from the starting configuration until
the first configuration in state $j_\forall$, the second block is the
sequence starting from this configuration until the first following
configuration in state $j_\exists$, and so on. We can assume that the first
nondeterministic configuration is a jump configuration and thus has
state~$j_\exists$.

Define the machine $\A'$ as follows. On an input $x$ of length at least $2$,
$\A'$ simulates $\A$ on $x$ and additionally stores the parity of (the number
of) the current block of the run. In an odd (even) block $\A'$ replaces
$\A$'s existential (universal) guesses of a bit $b$ by existential
(universal) jumps. Namely, it first computes the number $m$ of the cell
currently read on the input tape, then performs an existential (universal)
jump, then computes the parity $b$ of the cell it jumps to, then moves the
input head back to cell $m$ and then continues the run of $\A$ with guessed
bit~$b$ (i.\,e.\ in state $c_b$).

Then $\A'$ accepts $Q$ and every run of $\A'$ on $x$ does not have any
configuration with state~$c_\exists$ ($c_\forall$) in an odd (even) block.
More precisely, and assuming $t=3$ for notational simplicity: every run of
$\A'$ on $x$ has a nondeterminism type which is a prefix of a word in
\begin{equation}\label{eq:simtype}
j_\exists {\{c_\forall, j_\exists\}}^{\le f(\kappa(x))} j_\forall {\{c_\exists, j_\forall\}}^{\le f(\kappa(x))}
j_\exists {\{c_\forall, j_\exists\}}^{\le f(\kappa(x))}.
\end{equation}

Since $\kappa$ is computable in logarithmic space, the number $k:=
f(\kappa(x))$ can be computed in parameterized logarithmic space. Define a
machine $\A''$ which on $x$ first computes $k:= f(\kappa(x))$ and then
simulates $\A'$ keeping record of the nondeterminism type of the sofar
simulated run. Moreover, $\A''$ uses the record to do appropriate dummy
nondeterministic steps to ensure its nondeterminism type to be
\[
j_\exists {(c_\forall j_\exists)}^{2k}c_\forall j_\forall {(c_\exists j_\forall)}^{2k} c_\exists
j_\exists {(c_\forall j_\exists)}^{2k}c_\forall. \benda
\]

\section{Model-checking problems and the tree hierarchy}\label{sec:modcheck}

In Section~\ref{sec:comp} we prove Theorem~\ref{thm:completeness}~(1) as
Theorem~\ref{thm:mccomplete}, and draw some corollaries to its proof in
Section~\ref{sec:cors}. In particular, Theorem~\ref{thm:completeness}~(2) is
proved as Corollary~\ref{cor:focomplete}, and the collapse theorem announced
in the introduction is proved as Corollary~\ref{cor:collapse}. In
Section~\ref{sec:function} we prove that the completeness results stay true
when function symbols are allowed.

\subsection{Completeness results}\label{sec:comp}
It is easy to see that $\pmc(\cl{FO}^s)\in\paraL$ for $s=1$. We prove
completeness results for~$s\ge 2$.

\begin{thm}\label{thm:mccomplete}
Let $t\ge 1$ and $s\ge 2$. Then $\pmc(\Sigma^s_t)$ is complete for
$\cl{Tree}[t]$ under pl-reductions.
\end{thm}

\begin{proof}
(Containment) We first show that $\pmc(\Sigma^s_t)\in \cl{Tree}[t]$. This is
done by a straightforward algorithm $\A$ as follows. At any moment it keeps
in memory a formula $\psi$ and an assignment~$\alpha$ to its free variables.
For simplicity we assume that all negation symbols of $\psi$ appear only in
front of atoms.

In case $\psi$ is an atom or a negated atom, $\A$ accepts if $\alpha$
satisfies $\psi$ in $\str A$ and rejects otherwise. If $\psi$ is a
disjunction $(\psi_0\vee \psi_1)$ \big(conjunction $(\psi_0\wedge
\psi_1)$\big), the algorithm $\A$ existentially (universally) guesses a bit
$b$ and recurses replacing $\psi$ by $\psi_b$ and $\alpha$ by its restriction
to the free variables in $\psi_b$. If $\psi$ is $\exists x\psi_0$
(resp.~$\forall x\psi_0$), then $\A$ makes an existential (resp.\ universal)
jump to guess $a\in A$ and recurses replacing $\psi$ by $\psi_0$ and $\alpha$
by the assignment which extends~$\alpha$ mapping $x$ to $a$. Here we assume
$x$ occurs freely in $\psi_0$ --- otherwise $\mathbb A$ simply recurses on
$\psi_0$ with $\alpha$ unchanged.

When started on a $\Sigma_t^s$-sentence $\varphi$ and the empty assignment,
the formulas occurring during the recursion are subformulas of $\varphi$ and
thus contain as most $s$ variables, so each of the assignments computed
during the recursion can be stored in space roughly $s\cdot \log |A|$.

\medskip
\noindent (Hardness) We now show that $\pmc(\Sigma^2_t)$ is hard for
$\cl{Tree}[t]$ under pl-reductions. Given a problem decided by a
$t$-alternating machine $\B$ in small space, we construct a structure akin to
the configuration graph of $\B$. Its universe consists of all small \emph{nondeterministic} configurations of~$\B$. We draw a directed edge from one
such configuration to another if there exists a deterministic computation of
$\B$ leading from the first configuration to the second. We then express
acceptance by a short (parameter bounded) sentence which is constructed in
two steps: first we give a direct and intuitive construction using function
symbols. Then, in a second step, we show how to eliminate these functions
symbols. Details follow.

Let $(Q,\kappa)\in \cl{Tree}[t]$ be given and choose a computable $f$ and a
$t$-alternating machine $\B$ with $f\circ \kappa$ jumps and~$f\circ\kappa$
bits such that $\B$ accepts $Q$ and runs in space $f(\kappa(x))+O(\log|x|)$.

Given $x\in {\{0,1\}}^*$ compute an upper bound $s= f(\kappa(x))+ O(\log|x|)$
on the space needed by $\B$ on $x$. Since $\kappa$ is computable in
logarithmic space, such a number $s$ can be computed in parameterized
logarithmic space. We can assume that $\B$ on $x$ always halts after at most
$m= 2^{f(\kappa(x))}\cdot |x|^{O(1)}$ steps. Note the binary representation
of $m$ can be computed in parameterized logarithmic space.

For two space $s$ configurations $c$, $c'$ of $\B$ on $x$, we say that $\B$
\emph{reaches $c'$ from~$c$} if there is a computation of~$\B$ leading from
$c$ to $c'$ of length at most $m$ that does neither pass through a
nondeterministic configuration nor through a configuration of space larger
than~$s$. In particular, $c$ cannot be nondeterministic but $c'$ can. We
assume that $\B$ reaches a nondeterministic configuration from the initial
configuration, i.\,e.\ the computation of $\B$ on $x$ is not deterministic.

We define a structure $\str A$ whose universe $A$ comprises all \big(length
$O(s)$ binary codes of\big) nondeterministic space $s$ configurations of $\B$
on $x$. It interprets a binary relation symbol $E$, unary function symbols
$s_0, s_1$ and unary relation symbols $S, F, J_\exists, J_\forall, C_\exists,
C_\forall$ as follows.

A pair $(c,c')\in A^2$ is in $E^{\str A}$ if there exists a successor
configuration~$c''$ of~$c$ such that $\B$ reaches~$c'$ from~$c''$. The
relation symbol $S$ is interpreted by $S^{\str A}=\{c_{\textrm{first}}\}$
where $c_{\textrm{first}}$ is the (unique) first configuration in $A$ reached
by~$\B$ from the initial configuration of~$\B$ on $x$. The relation symbols
$J_\exists, J_\forall, C_\exists$ and $C_\forall$ are interpreted by the sets
of configurations in $A$ with states $j_\exists, j_\forall, c_\exists$ and
$c_\forall$ respectively. Obviously these sets partition $A$.

The relation symbol $F$ is interpreted by the set $F^{\str A}$ of those $c\in
A$ such that one of the following holds:
\begin{enumerate}
\item[--] $c\in C^{\str A}_\exists\cup J^{\str A}_\exists$ and $\B$ reaches
    a space $s$ accepting halting configuration from at least one successor
    configuration of $c$;

\item[--] $c\in C^{\str A}_\forall\cup J^{\str A}_\forall$ and $\B$ reaches
    a space $s$ accepting halting configuration from all successor
    configurations of $c$.
\end{enumerate}

\noindent
Finally, the function symbols $s_0$ and $s_1$ are interpreted by any
functions $s_0^{\str A}, s_1^{\str A}: A\to A$ such that for every $c\in
C_\exists^{\str A}\cup C_\forall^{\str A}$ with $\{d\mid (c,d)\in E^{\str
A}\}\neq \emptyset$ we have:
\[
\big\{s_0^{\str A}(c), s_1^{\str A}(c)\big\}
 = \big\{d\in A\bigmid (c,d)\in E^{\str A}\big\}.
\]

It is easy to check that $\str A$ is computable from $x$ in parameterized
logarithmic space. For example, to check whether a given pair $(c,c')\in A^2$
is in $E^{\str A}$ we simulate $\B$ starting from $c$ for at most $m$ steps;
if the simulation wants to visit a configuration of space larger than $s$ or a
nondeterministic configuration $\ne c'$, then we stop the simulation and
reject.

For a word $w$ of length $|w|\ge 1$ over the alphabet $\{j_\exists,
j_\forall, c_\exists, c_\forall\}$ we define a formula~$\varphi_w(x)$ with
(free or bound) variables $x,y$ as follows. We proceed by induction on the
length $|w|$.

If $|w|=1$, define $\varphi_w(x):= F(x)$. For $|w|\ge 1$ define:
\begin{align*}
\varphi_{c_\forall w}(x) &:= C_\forall (x)\wedge \big(\varphi_w(s_0(x))\wedge \varphi_w(s_1(x))\big), \\
\varphi_{c_\exists w}(x) &:= C_\exists(x)\wedge \big(\varphi_w(s_0(x))\vee \varphi_w(s_1(x))\big), \\
\varphi_{j_\exists w}(x) &:= J_\exists (x)\wedge \exists y \big(E(x,y)\wedge \exists x(x=y \wedge \varphi_w(x))\big), \\
\varphi_{j_\forall w}(x) &:= J_\forall( x)\wedge \forall y \big(\neg E(x,y)\vee \forall x(\neg x=y \vee\varphi_w(x))\big).
\end{align*}

Let $|w|\ge 1$ and assume that $c\in A$ is a configuration such that every
run of $\B$ on $x$ starting at $c$ has nondeterminism type $w$
and consists of space $s$ configurations;
then
\begin{eqnarray}\label{eq:phiw}
\textup{$c$ is accepting} & \Longleftrightarrow & \str A\models \varphi_w(c).
\end{eqnarray}
This follows by a straightforward induction on $|w|$. Now we look for $\str
A'$ and $\varphi'_w\in\Sigma^2_t$
with this property but in a relational vocabulary.

By the Normalization Lemma~\ref{lem:normal} we can assume that all runs of
$\B$ on $x$ have nondeterminism type $w$ of the form~\eqref{eq:type}. For
such a $w$ we observe that $\varphi_w(x)$ has the required quantifier
structure: in the notation of the next Section~\ref{sec:function} it is in
$\textup{func-}\Sigma_t^2$, i.\,e.\ the class of formulas defined as
$\Sigma^2_t$ but allowing function symbols.

Furthermore,~$\varphi_w(x)$ does not contain nested terms, in fact, all its
atomic subformulas containing some function symbol are of the form
$E(s_b(x),y)$, $J_\exists(s_b(x))$, or $J_\forall(s_b(x))$. For $b\in\{0,1\}$
we introduce binary relation symbols $E_b$ and unary relation symbols
$J_{\forall b}$ and~$J_{\exists b}$, and then replace the atomic subformulas
$E(s_b(x),y)$, $J_\exists(s_b(x))$, $J_\forall(s_b(x))$ in $\varphi_{w}(x)$
by $E_b(x,y)$, $J_{\exists b}(x)$, $J_{\forall b}(x)$ respectively. This
defines the formula $\varphi'_w(x)$. Since
$\varphi_w(x)\in\textup{func-}\Sigma_t^2$, we have $\varphi'_w(x)\in
\Sigma_t^2$.

To define $\str A'$ we expand $\str A$ by interpreting the new symbols by:
\begin{align*}
E^{\str A'}_b &:= \big\{(c,d)\bigmid (s^{\str A}_b(c),d)\in E^{\str A}\big\}, \\
J^{\str A'}_{\exists b} &:= \big\{c\bigmid s^{\str A}_b(c)\in J_\exists^{\str A}\big\},\\
J^{\str A'}_{\forall b} &:= \big\{c\bigmid s^{\str A}_b(c)\in J_\forall^{\str A}\big\}.
\end{align*}
We have for all $c\in A$:
\begin{eqnarray*}
\str A\models \varphi_w(c) & \Longleftrightarrow & \str A'\models \varphi'_w(c).
\end{eqnarray*}
As the assumption of~\eqref{eq:phiw} is satisfied for $c_{\mathrm{first}}$,
and $c_{\mathrm{first}}$ is accepting if and only if $\B$ accepts~$x$, that
is, if and only if $x\in Q$, we get
\begin{eqnarray*}
x\in Q & \Longleftrightarrow & \str A'\models\varphi'_w(c_{\mathrm{first}}).
\end{eqnarray*}
Setting $\psi:= \exists x(S(x)\wedge \varphi'_w(x))$ we get a reduction as
desired by mapping $x$ to $(\psi,\str A')$.
\end{proof}

\subsection{Corollaries}\label{sec:cors}

The easy elimination of function symbols in the proof of Theorem~\ref{thm:mccomplete} rests on the fact that it is applied only to formulas
where function symbols are not nested. We shall treat the general case in the
next subsection. Before that we draw some consequences of the last theorem
and its proof.

\begin{cor}\label{cor:collapse}
Let $t'> t\ge 1$. If $\cl{Tree}[t]$ is closed under complementation, then
\[
\cl{Tree}[t']= \cl{Tree}[t].
\]
\end{cor}

\begin{proof}
It is sufficient to show this for $t'= t+1$. Assume $\cl{Tree}[t]$ is closed
under complementation. By the previous theorem it suffices to show
$\pmc(\Sigma^2_{t+1})\in \cl{Tree}[t]$, and we know $\pmc(\Sigma^2_{t})\in
\cl{Tree}[t]$. By assumption we find a computable $f$ and a $t$-alternating
machine~$\A'$ with $f\circ\kappa$ jumps and $f\circ\kappa$ bits that runs in
parameterized logarithmic space and accepts the complement of
$\pmc(\Sigma^2_{t})$.

We describe a machine $\B$ accepting $\pmc(\Sigma^2_{t+1})$. On input
$(\varphi,\str A)$ with a sentence $\varphi\in\Sigma^2_t$ we first eliminate
all void universal quantifiers, i.\,e.\ replace all subformulas $\forall
x\psi$ by~$\psi$ whenever~$x$ does not appear free in $\psi$. The machine
$\B$ starts simulating the recursive algorithm $\A$ for
$\pmc(\Sigma^2_{t+1})$ described in the previous proof. The simulation is
stopped when~$\A$ recurses to a subformula of $\varphi$ of the form $\forall
x\psi$. In this case, $\forall x\psi$ has at most one free variable $y$ and
the current assignment $\alpha$ maps it, say, to $a\in A$. The machine $\B$
simulates~$\A'$ on $(\chi,\str A')$ where $\str A'$ expands $\str A$
interpreting a new unary relation symbol $C$ by $\{a\}$ and where $\chi$ is a
$\Sigma^2_t$-sentence logically equivalent to $\neg \forall xy(C(y)\to\psi)$.

It is clear that $\B$ accepts~$\pmc(\Sigma^2_{t+1})$. Observe that $\B$ does
not make universal jumps before it starts simulating $\A'$. Hence, since
$\A'$ is $t$-alternating, so is $\B$. The number of jumps and guesses in a
run of $\B$ before the simulation of $\A'$ is clearly bounded by $|\varphi|$,
the parameter. Furthermore, $\B$ can be implemented in parameterized
logarithmic space: for the simulation of $\A'$ it stores $(\psi,a)$ and
relies on the implicit pl-computability of $(\chi,\str A')$ from~$(\psi,a,\str A)$. Thus, $\B$ witnesses that $\pmc(\Sigma^2_{t+1})\in
\TREE[t]$.
\end{proof}

\begin{cor}\label{cor:focomplete}
Let $s\ge 2$. Then $\pmc(\cl{FO}^s)$ is complete for $\cl{Tree}[*]$ under
pl-reductions.
\end{cor}

\begin{proof}
That $\pmc(\cl{FO}^s)\in \TREE[*]$ can be seen as in the proof of
Theorem~\ref{thm:mccomplete}. Hardness of $\pmc(\cl{FO}^2)$ also follows as
in this proof, but the argument is actually simpler: let
$(Q,\kappa)\in\TREE[*]$ and choose a machine accepting it according to the
Normalization Lemma~\ref{lem:normal}~(2). Observe that the formula
$\varphi_w$ does not contain the function symbols $s_0,s_1$ for $w$ as
in~\eqref{eq:typestar}. Hence the reduction can simply map $x\in{\{0,1\}}^*$
to $(\exists x(S(x)\wedge\varphi_w(x)),\str A'')$ where $w= {(j_\exists
j_\forall)}^{f(\kappa(x))}$ and $\str A''$ is obtained from $\str A$ by
forgetting the interpretations of $s_0,s_1$. It follows from~\eqref{eq:phiw}
that this defines a pl-reduction from $(Q,\kappa)$ to $\pmc(\cl{FO}^2)$.
\end{proof}

\begin{cor}\label{cor:treefpt}
$\TREE[*]\subseteq\cl{FPT}$.
\end{cor}

\begin{proof}
By Corollary~\ref{cor:focomplete} and the fact that
$\textsc{mc}(\FO^2)$ is in $\cl{P}$. 
\end{proof}

The introduction mentioned the result that the classical problem
$\textsc{mc}(\FO^s)$ is P-complete for~$s\ge 2$. As a further corollary we
get that the parameterized analogue of this completeness result is likely
false:

\begin{cor}\label{cor:compl}
Unless $\cl{P}=\cl{L}$, there is no $s\in\N$ such that $\pmc(\FO^s)$ is
$\cl{FPT}$-complete under pl-reductions.
\end{cor}

\begin{proof}
If $\pmc(\FO^s)$ is FPT-complete, then (we can assume $s\ge 2$ and thus)
$\TREE[*]=\cl{FPT}$ by Corollary~\ref{cor:focomplete}. This implies
$\cl{P}=\cl{L}$ by Proposition~\ref{prop:nlvstreestar}. \end{proof}

\subsection{Function symbols}\label{sec:function}
Let $\textup{func-}\Sigma_t$ be defined as $\Sigma_t$ except that function
symbols are allowed. It is not hard to show (see~e.g.~\cite[Example~8.55]{flumgrohebuch}) that $\pmc(\textup{func-}\Sigma_t)$ is
equivalent to $\pmc(\Sigma_t)$ under fpt-reductions (even pl-reductions). An
analogous statement for the model-checking problems characterizing the
classes of the W-hierarchy is not known. In fact, allowing function symbols
gives problems complete for the presumably larger classes of the
W$^{\mathrm{func}}$-hierarchy. We refer to~\cite{chenmachine} for more
information.

\medskip
Let $\textup{func-}\Sigma_t^s$ and $\textup{func-}\FO^s$ be defined as
$\Sigma^s_t$ and $\FO^s$, respectively, except that function symbols are
allowed.

\begin{thm}\label{thm:func}
Let $s\ge 2$ and $t\ge 1$. Then
\begin{enumerate}
\item $\pmc(\Sigma_t^s)\equiv_\pl \pmc(\textup{func-}\Sigma_t^s)$.

\item $\pmc(\FO^s)\equiv_\pl \pmc(\textup{func-}\FO^s)$.
\end{enumerate}
\end{thm}

\begin{proof}
The second statement will be an easy corollary to the proof of the first. We
first give a rough, informal sketch of this proof. The idea is, as usual, to
replace functions by their graphs and translate formulas to the resulting
relational vocabulary using formulas $\myvalue_t(x,\bar x)$ expressing that
$x$ is the value of the term $t$ at $\bar x$. This formula implements the
bottom-up evaluation of $t$ at $\bar x$ in a straightforward way. However, to
do so, the formula needs to quantify over tuples of intermediate values. To
do this using a single variable we extend the structure to contain tuples of
appropriate lengths. This enables us to write $\myvalue_t(x,\bar x)$ and the
whole translation with only a constant overhead of new variables. To preserve
the quantifier alternation rank we write two versions of $\myvalue_t(x,\bar{x})$, an existential and a universal one. Details follow.

\medskip

To prove statement (1) it suffices, by Theorem~\ref{thm:mccomplete}, to show
\[
\pmc(\textup{func-}\Sigma_t^s)\le_\pl \pmc(\Sigma_t^{s+3}).
\]

Let $\str{A}$ be a structure of a vocabulary $\tau$ with $|A|\ge 2$. We
define a relational vocabulary~$\tau'$ which depends on $\tau$ only, and a
$\tau'$-structure $\str{A}'$. The universe $A'$ is the union
\begin{itemize}
\item[--] of $A$;

\item[--] for each relation symbol $R \in \tau$ of arity $r$, of the set
    \begin{align*}
    \big\{(a_1, \ldots, a_i) \bigmid & \text{$2 \le i \le r$
     and $(a_1, \ldots, a_i, a_{i+1}, \ldots, a_r) \in R^{\str{A}}$} 
    \text{ for some $a_{i+1}, \ldots, a_r \in A$}\big\},
    \end{align*}
    i.\,e.\ the set of all `partial' tuples that can be extended to some
    tuple in $R^{\str{A}}$; note that the size of this set is bounded by
    $(r-1) \cdot |R^{\str{A}}|$;

\item[--] for each function symbol $f\in \tau$ of arity $r$, of the set
    \[\textstyle
    \bigcup_{2 \le i \le r} A^i.
    \]
\end{itemize}
We identify $A^1$ with $A$ and it is therefore that the union $\bigcup_{2
\le i \le r} A^i$ starts at $i= 2$. Recall we assumed $|A| \ge 2$, so
$\bigcup_{2 \le i \le r} A^i$ has size at most $|A|^{r+1}$. Altogether,
$|A'|\le |\str{A}|^2$.

Now we define the vocabulary $\tau'$ and the $\tau'$-structure $\str{A}'$ in
parallel:
\begin{itemize}

\item[--] $\tau'$ contains a unary relation symbol $U$ interpreted by the
    original universe of $\str A$, that is,
    \[
    U^{\str{A}'}:= A.
    \]

\item[--] For every constant symbol $c\in \tau$, the vocabulary $\tau'$
    contains a unary relation symbol $U_c$ which is interpreted as
    \[
    U^{\str{A}'}_c:= \big\{c^{\str{A}}\big\}.
    \]

\item[--] For every $r$-ary relation symbol $R \in \tau$, the vocabulary
    $\tau'$ contains a unary relation symbol~$U_R$ and we set
    \[
    U^{\str{A}'}_R:= \big\{b\in A'\bigmid
     \text{$b=(a_1, \ldots, a_r)$ for some $(a_1, \ldots, a_r)
      \in R^{\str{A}}$} \big\}.
    \]

\item[-] For every $r$-ary function symbol $f \in \tau$, the vocabulary
    $\tau'$ contains a binary relation symbol~$F_f$ and we define
    \[
    F^{\str{A}'}_f:= \big\{((a_1, \ldots, a_r),a)\in A'\bigmid
      \text{$(a_1, \ldots, a_r)\in A^r$ and $f^{\str{A}}(a_1, \ldots, a_r)= a$} \big\}.
    \]

\item[--] $\tau'$ contains a ternary relation symbol $R_e$; the `tuple
    \emph{e}xtending relation' $R^{\str{A}'}_e$ is defined by
    \begin{align*}
    R^{\str{A}'}_e:= \big\{(b,a,b')\in A'\times A\times A'\bigmid
    & \text{there are $i\ge 1$ and $a_1, \ldots, a_i\in A$ such that}\\
    & b= (a_1, \ldots, a_i)\in A'\text{ and } b'= (a_1, \ldots, a_i, a)\in A' \big\},
    \end{align*}
\end{itemize}

\noindent
Now, for any sentence $\varphi \in \textup{func-}\Sigma_t^s$, we construct
two sentences $\trans^{\exists}_{\varphi}$ and $\trans^{\forall}_{\varphi}$
such that
\begin{eqnarray}\label{eqn:translation}\nonumber
\str{A} \models \varphi &\iff &\str{A}' \models \trans^{\exists}_{\varphi} \\
 & \iff & \str{A}' \models \trans^{\forall}_{\varphi}.
\end{eqnarray}

We start by defining, for every term $m(\bar x)$ with $\bar x= x_1, \ldots,
x_s$, two formulas
\begin{eqnarray*}
\myvalue^{\exists}_m(x,\bar x)
 & \text{and} &
\myvalue^{\forall}_m(x, \bar x)
\end{eqnarray*}
which respectively are in $\Sigma_1^{s+3}$ and $\Pi_1^{s+3}$ up to logical
equivalence. Furthermore, for every $a\in A'$ and every $a_1, \ldots, a_s
\in A$ it holds that
\begin{align*}
a= m^{\str{A}}(a_1, \ldots, a_s)
 \iff & \str{A}' \models \myvalue^{\exists}_m(a, a_1, \ldots, a_s) \\
 \iff & \str{A}' \models \myvalue^{\forall}_m(a, a_1, \ldots, a_s).
\end{align*}
If $m$ is a variable $x_i$, then $\myvalue^{\exists}_m:=
\myvalue^{\forall}_m:= x=x_i$. If $m$ is a constant $c$, then
\[\myvalue^{\exists}_m:= \myvalue^{\forall}_m:= U_c(x).\]
If $m$ is the
composed term $f(m_1, \ldots, m_r)$, then
\begin{align*}
\myvalue^{\exists}_m(x, \bar x) & := \exists y \Big(
 \exists x \big(x=y \wedge \tuple_{m_1, \ldots, m_r}(x, \bar x)\big)
 \wedge F_f(y,x)\Big), \\
\myvalue^{\forall}_m(x, \bar x) & := \forall y \Big(
 \neg \exists x \big(x=y \wedge\tuple_{m_1, \ldots, m_r}(x, \bar x)\big)
 \vee F_f(y,x)\Big),
\end{align*}
where $\tuple_{m_1, \ldots, m_r}$ is defined inductively on $r$ as follows.
\begin{align*}
\tuple_{m_1}(x, \bar x) &:= \myvalue^{\exists}_{m_1}(x, \bar x), \\
\tuple_{m_1, \ldots, m_{i+1}}(x, \bar x)
 &:= \exists y \exists z\Big( R_e(y,z,x)\wedge
  \exists x\big(x=y \wedge \tuple_{m_1, \ldots, m_i}(x, \bar x)\big) \\
 & \hspace{4.5cm} \wedge \exists x\big(x=z \wedge \myvalue^{\exists}_{m_i}(x, \bar x)\big) \Big).
\end{align*}
It is easy to verify that $\tuple_{m_1, \ldots, m_r}\in \Sigma^{s+3}_1$ and
for every $b\in A'$ and every $a_1, \ldots, a_s \in A$
\begin{eqnarray*}
b= \Big(m^{\str{A}}_1(a_1, \ldots, a_s), \ldots, m^{\str{A}}_r(a_1, \ldots, a_s)\Big)
 & \iff & \str{A}' \models \tuple_{m_1, \ldots, m_r}(b, a_1, \ldots, a_s).
\end{eqnarray*}

\medskip
For formulas $\varphi$ we define $\trans^{\exists}_{\varphi}$ and
$\trans^{\forall}_{\varphi}$ by induction as follows:
\begin{align*}
\trans^{\exists}_{m_1 = m_2} &:= \exists x \big(\myvalue^{\exists}_{m_1}(x, \bar x)
 \wedge \myvalue^{\exists}_{m_2}(x, \bar x)\big), \\
\trans^{\forall}_{m_1 = m_2} &:= \forall x
 \big(\neg \myvalue^{\exists}_{m_1}(x, \bar x) \vee \myvalue^{\forall}_{m_2}(x, \bar x)\big), \\
\trans^{\exists}_{Rm_1\ldots m_r} &:= \exists x \big(U_R(x) \wedge
 \tuple_{m_1, \ldots, m_r}(x, \bar x)\big), \\
\trans^{\forall}_{Rm_1\ldots m_r} &:= \forall x \big(\neg \tuple^{\exists}_{m_1, \ldots, m_r}(x, \bar x)
\vee U_R(x)\big).
\end{align*}
If $\varphi= \neg \psi$, then $\trans^{\exists}_{\varphi}:= \neg
\trans^{\forall}_{\psi}$ and $\trans^{\forall}_{\varphi}:= \neg
\trans^{\exists}_{\psi}$. If $\varphi= (\psi_1\vee \psi_2)$, then
$\trans^{\exists}_{\varphi}:= \big(\trans^{\exists}_{\psi_1}\vee
\trans^{\exists}_{\psi_2}\big)$ and $\trans^{\forall}_{\varphi}:=
\big(\trans^{\forall}_{\psi_1}\vee \trans^{\forall}_{\psi_2}\big)$. The case
for $(\psi_1 \wedge \psi_2)$ is similar. If $\varphi = \exists x_i \psi$, we
define
\begin{eqnarray*}
\trans^{\exists}_{\varphi} := \exists x_i \big(U_A(x_i) \wedge \trans^{\exists}_{\psi}\big)&
\text{and} & \trans^{\forall}(\varphi) := \exists x_i \big(U_A(x_i) \wedge \trans^{\forall}_{\psi}\big).
\end{eqnarray*}
Similarly, for $\varphi = \forall x \psi$,
\begin{eqnarray*}
\trans^{\exists}_{\varphi} := \forall x_i \big(\neg U_A(x_i) \vee \trans^{\exists}_{\psi}\big)
 & \text{and} &
\trans^{\forall}_{\varphi} := \forall x_i \big(\neg U_A(x_i) \vee \trans^{\forall}_{\psi}\big).
\end{eqnarray*}
It is routine to verify~\eqref{eqn:translation}.
%
Moreover, if $t$ is odd, then $\trans^{\exists}_{\varphi}$ is equivalent to
a $\Sigma^{s+3}_t$-sentence; and if $t$ is even, then
$\trans^{\forall}_{\varphi}$ is equivalent to a $\Sigma^{s+3}_t$-sentence.
For simplicity, we denote the corresponding $\Sigma^{s+3}_t$-sentence and
$\Sigma^{s+3}_t$-sentence by $\trans^{\exists}_{\varphi}$ and
$\trans^{\forall}_{\varphi}$ again. Therefore, for every structure $\str{A}$
and $\varphi\in \textup{func-}\Sigma_t^s$
\[
\big(\str{A}, \varphi\big) \mapsto
\begin{cases}
\big(\str{A}', \trans^{\exists}_{\varphi}\big) & \text{if $t$ is odd}, \\
\big(\str{A}', \trans^{\forall}_{\varphi}\big) & \text{if $t$ is even}
\end{cases}
\]
gives the desired pl-reduction from $\pmc(\textup{func-}\Sigma_t^s)$ to
$\pmc(\Sigma_t^{s+3})$. This finishes the proof of statement (1).

\medskip

The proof of statement (2) is now easy. By
Corollary~\ref{cor:focomplete}, it is enough to show
\[
\pmc(\textup{func-}\FO^s)\le_\pl \pmc(\FO^{s+3}).
\]
This is witnessed by the reduction mapping an instance $(\str A,\varphi)$ of
$\pmc(\textup{func-}\FO^s)$ to the instance $\big(\str{A}',
\trans^{\exists}_{\varphi}\big)$, defined as above.
\end{proof}


\section{PATH and optimality of Savitch's Theorem}

Savitch's Theorem is a milestone result linking nondeterministic space to
deterministic space.
\nprob{\stdipath}{A directed graph $\mathbf G$ and two vertices $s, t\in
G$}{Is there a (directed) path from $s$ to $t$ in $\mathbf G$?}

\begin{thm}[Savitch~\cite{sav}]\label{theo:savitch}
$\stdipath\in\cl{SPACE}(\log^2 n)$. In particular,
\[\cl{NL}\subseteq \cl{SPACE}(\log^2 n).\]
\end{thm}

The second statement follows from the first via the following proposition,
itself a direct consequence of the fact that \stdipath\ is complete for NL
under logarithmic space reductions (see
e.g.~\cite[Theorem~4.18]{arorabarak}).

\begin{prop}\label{prop:oed}
%
Let $s:\N\to \N$ and assume $\stdipath\in \cl{SPACE}(s)$. Then
\[
\cl{NL}\subseteq \cl{SPACE}\left(s\big(n^{O(1)}\big)+ \log n\right).
\]
\end{prop}

In this section we prove a stronger version of
Theorem~\ref{thm:improvesavitch} and Theorem~\ref{thm:lower}. Additionally,
we explain what a collapse of the parameterized classes $\PATH$ and $\paraL$
means in terms of classical complexity classes.

\subsection{Proof of Theorem~\ref{thm:improvesavitch}}

Recall we say \emph{Savitch's Theorem is optimal} if
\[\cl{NL}\not\subseteq\cl{SPACE}(o(\log^2n)).\]
In this subsection we prove:

\improvesavitch*

In fact, we shall prove something stronger, namely Theorem~\ref{thm:savitch}
below: its assumption for \emph{computable} $f$ is equivalent to
$\PATH\subseteq\paraL$ by Theorem~\ref{thm:pathcomplete}; its conclusion
implies that Savitch's Theorem is not optimal via Proposition~\ref{prop:oed}.
We shall use this stronger statement when proving Theorem~\ref{thm:lower} in
the next subsection.

\begin{thm}\label{thm:savitch}
Assume there is an algorithm deciding $p\textsc{-\STCon}$ that on instance
$(\str G,s,t,k)$ runs in space
\begin{equation}\label{eq:Aspace}
f(k)+ O(\log |G|)
\end{equation}
for some $f:\N \to \N$ (not necessarily computable). Then
$\stdipath\in\cl{SPACE}(o(\log^2 n))$.
\end{thm}

\begin{proof} Choose an algorithm $\A$ and a function $f$ according to the
assumption. Without loss of generality, assume that
$f(k)\ge k$ for every $k\in \N$. Let $\iota: \N\to \N$ be a non-decreasing
and unbounded function such that for all $n\in \N$
\begin{equation}\label{eq:iota}
f(\iota(n))\le \log n,
\end{equation}
and hence
\begin{equation}\label{eq:iotabound}
\iota(n)\le \log n.
\end{equation}
Note that we might not know how to compute $\iota(n)$.

\medskip
Now let $\mathbf G$ be a directed graph, $s,t \in G$, $n:= |G|$, and $k\ge
2$. We compute in space $O(\log k+ \log n)$ the \emph{minimum} $\ell:=
\ell(k) \in \N$ with
\begin{equation}\label{eq:kln}
k^{\ell} \ge n-1,
\end{equation}
which implies
\begin{equation}\label{eq:ell}
\ell\le O\left(\frac{\log n}{\log k}\right). 
\end{equation}
Then we define a sequence of directed graphs ${\left(\mathbf G^k_i\right)}_{i \in\{0,\ldots, \ell\}}$ with self-loops, as follows. For every $i\ge 0$ the
vertex set $G^k_i$ of $\str G_i^k$ is $G$,
the vertex set of $\str G$. There is an edge in $\mathbf G^k_i$ from a
vertex~$u$ to a vertex $v$ if and only if
there is a directed path from $u$ to $v$ in $\mathbf G$ of length at most
$k^i$. In particular, $E^{\mathbf G^k_0}$ is the reflexive closure of
$E^{\mathbf G}$. By~\eqref{eq:kln}
\begin{eqnarray}\label{eq:Gt}
\text{there is a path from $s$ to $t$ in $\mathbf G$}
 &\iff &
\text{there is an edge from $s$ to $t$ in $\mathbf G^k_{\ell}$}.
\end{eqnarray}
Furthermore, for every $i\in [\ell]$ and $u,v\in G^k_i= G^k_{i-1}= G$
\begin{align*}
\text{there is an edge} & \ \text{from $u$ to $v$ in $\mathbf G^k_i$} \\
 &\iff
\text{there is a path from $u$ to $v$ in $\str G^k_{i-1}$ of length at most $k$}.
\end{align*}
This can be decided by the following \emph{recursive} algorithm:

\begin{center}
\fbox{
\begin{minipage}[t]{13cm}
Algorithm $\C$ \\
\textit{input:} a directed graph $\str G$, $k, i\in \N$, and $u,v \in G$\\
\textit{output:} decide whether there is an edge in $\str G^k_i$ from $u$ to
$v$.
\begin{algorithm}

\im0 \IF\ $i=0$ \THEN\ output whether \big($u=v$ or $(u,v)\in E^{\str
G}$\big) and return

\im0 simulate $\A$ on $\big(\str G^k_{i-1}, u, v, k\big)$

\im1 \IF\ in the simulation of $\A$ queries ``$(u',v') \in E^{\str
G^k_{i-1}}$?''

\im1 \quad \THEN\ call $\C(\mathbf G, k, i-1, u', v')$.

\end{algorithm}
\end{minipage}
}
\end{center}

\medskip
For every $k\ge 2$ let $\C^k$ be the algorithm which on every directed graph
$\str G$ and $s,t\in G$ first computes $\ell= \ell(k)$ as in~\eqref{eq:kln}
and then simulates $\C(\mathbf G, k, \ell, s, t)$. Thus, $\C^k$ decides
whether there is a path from $s$ to $t$ in $\mathbf G$ by~\eqref{eq:Gt}. We
analyse its space complexity. First, the depth of the recursion tree is
$\ell$, as $\C^k$ recurses on $i=\ell, \ell-1, \ldots, 0$. As usual, $\C^k$
has to maintain a stack of intermediate configurations for the simulations
of
\[
\A(\mathbf G^k_{\ell}, \_, \_, k), \A(\mathbf G^k_{\ell-1}, \_, \_, k), \ldots,
 \A(\mathbf G^k_0, \_, \_, k).
\]
For the simulation of each $\A(\mathbf G^k_i, \_, \_, k)$, the size of the
configuration is linearly bounded by $f(k)+ O(\log n)$ because
of~\eqref{eq:Aspace}. Therefore, the total space required is
\begin{equation}\label{eq:totalspaceCk}
O\Big(\log k+ \log n+ \ell\cdot \big(f(k)+\log n\big)\Big)
 \le O\left(\log k+ \frac{f(k)\cdot \log n+ \log^2 n}{\log k}\right) 
\end{equation}
by~\eqref{eq:ell}. As a consequence, in case $k= \iota(n)$,~\eqref{eq:iota}
and~\eqref{eq:iotabound} imply that~\eqref{eq:totalspaceCk} is bounded
by~$o(\log^2 n)$. So if $\iota(n)$ would be computable, we could replace it
with a logspace computable function, and then the result would follow. In
particular, under the assumption $\PATH=\paraL$ of
Theorem~\ref{thm:improvesavitch}, we can assume $f$ is computable, and hence
find a computable $\iota$.

\medskip
In order to circumvent the possible uncomputability of $\iota(n)$ we adopt
the strategy underlying Levin's optimal inverters~\cite{levin, chefluBSL}.
Namely, we simulate all the algorithms $\C^2, \C^3, \ldots$ in a diagonal
fashion, while slowly increasing the allowed space.
\begin{center}
\fbox{
\begin{minipage}[t]{13cm}
Algorithm $\mathbb S$ \\
\textit{input:} a graph $\str G$ and $s,t \in G$\\
\textit{output:} decide whether there is a path in $\str G$ from $s$ to $t$.
\begin{algorithm}

\im0 $S\gets 2$

\im0 \FORALL\ $i=2$ \TO\ $S$ \DO%

\im1 simulate $\C^i$ on $\big(\str G, s, t\big)$ in \emph{space $S$}

\im1 \IF\ the simulation accepts or rejects in space $S$

\im2 \THEN\ accept or reject accordingly

\im0 $S\gets S+1$

\im0 goto 2.
\end{algorithm}
\end{minipage}
}
\end{center}
Clearly, $\mathbb S$ decides \stdipath. We prove that its space complexity
is $o(\log^2 n)$ on every input graph~$\str G$ with $n:= |G|$. To that end,
let $k:= \iota(n)$ and $s^*$ be the space needed by $\C^k(\str G, s, t)$. As
argued before, we have $s^*\le o(\log^2 n)$. Observe that $\mathbb S$ must
halt no later than $S$ reaches the value $\max(\iota(n), s^*)$, which is
again bounded by $o(\log^2 n)$ due to~\eqref{eq:iotabound}. This concludes
the proof.
\end{proof}

\subsection{Proof of Theorem~\ref{thm:lower}}

For the reader's convenience, we repeat the statement of the theorem:

\thmlower*

Note that neither $f$ nor the function hidden in the $o(\ldots)$-notation is
assumed to be computable. It is here where our stronger version
Theorem~\ref{thm:savitch} of Theorem~\ref{thm:improvesavitch} becomes useful.

\begin{proof} Assume $\pmc(\Sigma_1^2)$ is decidable in space
$o(f(|\varphi|)\cdot \log |\str A|)$ for some function $f:\N\to\N$, i.\,e.\
space $g\big(f(|\varphi|)\cdot \log |\str A|)$ for some function $g:\N
\to \N$ with
\begin{equation}\label{eq:limg}
\lim_{m\to \infty} \frac{g(m)}{m}=0.
\end{equation}
By Theorem~\ref{thm:savitch} it suffices to show that there exist an
arbitrary function $h:\N\to\N$ and an algorithm deciding $p$-$\STCon$ that on
an instance $(\str G,s,t,k)$ runs in space
\begin{equation}\label{eq:h}
h(k)+O(\log |G|).
\end{equation}

Example~\ref{ex:fostconrel} defines a pl-reduction from $p$-$\STCon$ to
$\pmc(\Sigma_1^2)$ that maps an instance $(\str G,s,t,k)$ of $p$-$\STCon$ to
an instance $(\varphi,\str A)$ of $\pmc(\Sigma_1^2)$ such that $|\str A|\le
O(|\str G|)$ and $|\varphi| \le c\cdot k$ for some constant $c\in \N$. Then,
by the assumption and this reduction, there is an algorithm $\A$ which
decides $p\textsc{-\STCon}$ in space
\[
g\big(f(c\cdot k) \cdot \log |\str G|\big),
\]
Here, we assume without loss of generality that $f$ is non-decreasing, so
$f(|\varphi|)\le f(c\cdot k)$. By~\eqref{eq:limg}, for every $\ell\in \N$
there is $m_{\ell}\in \N$ such that for every $m\ge m_{\ell}$ we have
\[
\frac{g(m)}{m}\le \frac{1}{\ell}.
\]
Thus, for $\ell:= f(c\cdot k)$ and $m:= f(c\cdot k) \cdot \log |\str G|$
\begin{align*}
g\big(f(c\cdot k) \cdot \log |\str G|\big) = g\big(m\big)
   \le \max_{m< m_{\ell}} g(m)+ \frac{m}{\ell}= \max_{m< m_{\ell}} g(m) + \log |\str G|.
\end{align*}
Note $\max_{m< m_{\ell}} g(m)$ only depends on $\ell$, and hence only on the
parameter $k$. Consequently the space required by $\A$ can be bounded
by~\eqref{eq:h} for an appropriate function $h:\N \to\N$.
\end{proof}

\subsection{Bounded nondeterminism in logarithmic space}

We close this section showing that the collapse of the parameterized classes
$\paraL$ and $\PATH$ can be characterized as a collapse of classical classes L and NL
restricted to `arbitrarily few but non-trivially many' nondeterministic bits.

\begin{defi}
Let $c: \N\to \N$ be a function. The class $\textup{NL}[c]$ contains all
classical problems~$Q$ that are accepted by some nondeterministic Turing
machine which uses $c(|x|)$ many nondeterministic bits and runs in
logarithmic space.
\end{defi}

\begin{prop}\label{prop:nl}The following are equivalent.
\begin{enumerate}
\item $\paraL= \PATH$.

\item There exists a space-constructible\footnote{Recall, $c:\N\to \N$ is
    \emph{space-constructible} if $c(n)$ can be computed from $n$ in space
    $O(c(n))$.} function $c(n)\ge \omega(\log(n))$ such that
    $\textup{NL}[c]= \textup{L}$.
\end{enumerate}
\end{prop}

\begin{proof}
To see that (1) implies (2), assume $\paraL= \PATH$. Then there is a machine
$\A$ deciding $ p\textsc{-\STCon}$ which on an instance $(\str G,s,t,k)$ runs
in space $f(k)+ O(\log |G|)$ for some computable $f:\N\to\N$. We can assume
that $f$ is increasing and space-constructible (see~\cite[Lemma~1.35]{flumgrohebuch} for a similar construction).
Then there is an unbounded, logarithmic space computable $\iota: \N\to \N$
such that $f(\iota(n))\le \log n$ for all $n\in\N$. Then
\[
c(n):= \iota(n)\cdot \log n
\]
is space-constructible and $c(n)\ge \omega(\log(n))$. We claim that
$\textup{NL}[c]= \textup{L}$.

Let $Q\in \textup{NL}[c]$ be given. Choose a machine $\B$ accepting $Q$ that
on input $x$ uses at most $s= O(\log |x|)$ space and at most
$\iota(|x|)\cdot \log|x|$ many nondeterministic bits. We assume there is at
most one accepting space $s$ configuration $c_{\textit{acc}}$ that can
possibly appear in any run of $\B$ on~$x$.

The digraph $\str G$ has as vertices all space $s$ configurations of $\B$ on
$x$ and an edge from $u$ to~$v$ if there is a computation of $\B$ started at
$u$ leading to $v$ which uses at most $\log|x|$ many nondeterministic bits
and space at most $s$. Note this can be decided in logarithmic space by
simulating $\B$ exhaustively for all possible outcomes of guesses.

Since $\iota$ is logarithmic space computable, the instance $\big(\str G,
c_{\textit{start}}, c_{\textit{acc}}, \iota(|x|)\big)$ of $p$-$\STCon$ is
implicitly logarithmic space computable from $x$; here, $c_{\textit{start}}$
is the starting configuration of $\B$ on~$x$. This is a ``yes'' instance if
and only if $x\in Q$. Given this input, $\A$ needs space
\[
f(\iota(|x|))+O(\log|G|)\le O(\log|x|).
\]

\medskip
To see that (2) implies (1), assume $c(n)\ge\omega(\log(n))$ is
space-constructible and $\textup{NL}[c]= \textup{L}$. There is a logarithmic
space computable, non-decreasing and unbounded function $\iota: \N\to \N$
such that $c(n)\ge \iota(n)\cdot \lceil\log n\rceil$ for all~$n\in\N$. By
Theorem~\ref{thm:pathcomplete} it suffices to show that $p$-$\STCon$ can be
decided in parameterized logarithmic space.

Define the classical problem
\[
Q:=\big\{ (\str G,s,t,k)\in \STCon \mid k\le\iota(|G|)\big\}.
\]
Then $Q\in\textup{NL}[c]$ by a straightforward guess and check algorithm. By
assumption, $Q$ is decided by a logarithmic space algorithm~$\A$. Then we
solve $p$-$\STCon$ as follows using some arbitrary ``brute force''
(deterministic) algorithm $\B$ deciding $p$-$\STCon$. Given an instance
$(\str G,s,t,k)$ we check whether $k\le \iota(|G|)$; if this is the case, we
run $\A$ and otherwise $\B$. In the first case we consume only logarithmic
space. In the second case, the space is effectively bounded in $k$ because
the instance size is. Indeed, if $k> \iota(|G|)$, then $|G|< f(k)$, where
$f$ is a computable function such that $f\circ\iota(n)\ge n$ for all
$n\in\N$.
\end{proof}

\begin{rem}
There are similar characterizations of $\textup{W[P]}= \textup{FPT}$
in~\cite[Theorem~3.8]{fellows}, $\cl{EW[P]}= \cl{EPT}$
in~\cite[Theorem~4]{fgw}, and $\textup{BPFPT}= \textup{FPT}$
in~\cite[Theorem~5.2]{montoya}.
\end{rem}

We find it worthwhile to point out explicitly the following direct
corollary.

\begin{cor}\label{cor:sav}
If Savitch's Theorem is optimal, then $\cl{L}\ne \cl{NL}[c]$ for all
space-constructible functions $c(n)\ge\omega(\log n)$.
\end{cor}

\begin{proof}
By Proposition~\ref{prop:nl} and Theorem~\ref{thm:improvesavitch}.
\end{proof}

\section{A deterministic model-checker}

In this section we prove

\thmupper*

This claims for each $s\in\N$ the existence of an algorithm. Our algorithm is
going to be uniform in $s$, so we give a more general statement formulating
the space bound using the \emph{width}~$w(\varphi)$ of a formula $\varphi$.
This is the maximal number of free variables in some subformula of~$\varphi$.

Being interested in the dependence of the space complexity on the parameter
$|\varphi|$, we formulate all results as statements about the classical
model-checking problem $\textsc{mc}(\Phi)$ for various classes $\Phi$ of
first-order sentences. In this section we allow besides relation symbols
also constants in the vocabulary of formulas in $\Phi$. Abusing notation we
continue to use notation like $\Sigma_t$ and $\Pi_t$, now understanding that
the formulas may contain constants.

Besides $|\varphi|$ it is technically convenient to also consider another
size measure of $\varphi$, namely the number of subformulas $\|\varphi\|$ of
$\varphi$ counted with repetitions. In other words this is the number of
nodes in the syntax tree of $\varphi$. Formally, it is defined by a
recursion on the syntax:
\[
\begin{array}{rcll}
\|\varphi\|&:=&1&\textup{for atomic $\varphi$}\\
\|(\varphi * \psi)\| &:=&1+\|\varphi\|+\|\psi\| &\textup{for $*\in\{\wedge,\vee\}$}\\
\|\neg\varphi\| &:=&1+\|\varphi\|&\\
\|Qx \varphi\| &:=&1+\|\varphi\|&\textup{for $Q\in\{\forall,\exists\}$}
\end{array}
\]
Clearly we have $\|\varphi\|\le|\varphi|$. Hence, the following result, which
is the main result of this section, implies Theorem~\ref{thm:upper}.

\begin{thm}\label{thm:sigmat} For all $t\ge 1$ there is
 an algorithm deciding $\textsc{mc}(\Sigma_t)$ that on an
instance~$(\varphi,\str A)$ of $\textsc{mc}(\Sigma_t)$ runs in space
\begin{equation*}
O\big( \log \|\varphi\|\cdot w(\varphi)\cdot\log |A|+ \log\|\varphi\|\cdot\log|\varphi|+ \log |\str A|\big).
\end{equation*}
\end{thm}

We first give an intuitive outline of the proof. The heart of the argument is
the proof for the case $t=1$, the extension to $t\ge 1$ is straightforward.
The case $t=1$ is isolated as Proposition~\ref{prop:sigma1}, and proved by a
recursive divide and conquer approach. Namely, given an instance~$(\str
A,\varphi)$ of~$\textsc{mc}(\Sigma_1)$, our model-checker views $\varphi$ as
a tree and computes a node that splits the tree in a 1/3--2/3 fashion. More
precisely, it computes a subformula $\varphi_0(\bar y)$ of $\varphi$ of size
$\|\varphi_0(\bar y)\|$ between~1/3 and 2/3 of $\|\varphi\|$. Our
model-checker loops through all possible assignments $\bar b$ to the free
variables $\bar y$ of $\varphi_0(\bar y)$ and recurses to~$\varphi_0(\bar{b})$.
Returning from this call it recurses to the ``rest''
formula~$\varphi_1^{\bar b}$. Intuitively this is the formula obtained by
replacing~$\varphi_0(\bar b)$ by its truth value.

As long as the formulas in the recursion are large enough, they shrink in
each recursive step by a constant fraction. When the recursion reaches a
small formula it applies ``brute force''. Hence the recursion tree is of
depth $O(\log\|\varphi\|)$. Since the tuples $\bar b$ from the loops can be
stored in space $O(w(\varphi)\cdot \log|A|)$, this sketch should explain the
main term $\log \|\varphi\|\cdot w(\varphi)\cdot\log |A|$ in the final space
bound.

We first describe the ``brute force'' subroutine mentioned above, a folklore,
straightforwardly defined model-checker:

\begin{lem}\label{lem:oed}
There is an algorithm deciding $\textsc{mc}(\FO)$ that on an
instance $(\varphi,\str A)$ of $\textsc{mc}(\FO)$ runs in space
\begin{equation*}\label{eq:allowoed}
O\big( \|\varphi\|\cdot\log |A|+ \log |\varphi|+ \log |\str A|\big).
\end{equation*}
\end{lem}

\begin{proof}
We describe an algorithm $\B$ that decides the slightly more general
(classical) problem
\nnprob{a formula $\varphi=\varphi(\bar x)$, a structure $\str A$ and a
tuple $\bar a\in A^{|\bar x|}$}{$\str A\models \varphi(\bar a)$?}
\noindent in space $O(\|\varphi\|\cdot\log |A|+ \log n)$ on an instance
$(\varphi,\str A,\bar a)$ of length $n$. The algorithm~$\B$ implements a
straightforward recursion on the logical syntax of $\varphi$.

If $\varphi(\bar x)=(\chi(\bar x) \wedge \psi(\bar x))$, it calls
$\B(\chi(\bar x),\str A,\bar a)$. Upon returning from this call, $\B$ stores
its answer, i.\,e.\ the bit $b$ giving the truth value of $\str
A\models\chi(\bar a)$. Then $\B$ calls $\B(\psi(\bar x),\str A,\bar a)$.
Upon returning from this call with answer $b'$, $\B$ answers the bit $b\cdot
b'$.

If $\varphi(\bar x)=\exists x\chi(\bar x,x)$, then $\B$ loops over $b\in A$
and calls $\B(\psi(\bar x,x),\str A,\bar ab)$; it answers with the maximal
answer bit obtained.

The cases where $\varphi$ is a negation, a disjunction or starts with
$\forall$ are similarly explained.

If $\varphi$ is atomic, it has the form $t_1{=}t_2$ or $R(t_1,\ldots, t_r)$
where $R$ is an $r$-ary relation symbol and the $t_i$'s are constants or
variables. Assume the latter. Then $\B$ checks whether there exists
$j\in[|R^{\str A}|]$ such that for all $i\in[r]$ it holds that $t_i^{A}$
equals the $i$-th component of the $j$-th tuple in $R^{\str A}$.
Here,~$t_i^{A}$ is~$c^{\str A}$ if~$t_i$ is a constant $c$; otherwise $t_i$
is a free variable in $\varphi$ and~$t_i^A$ is the corresponding component
of~$\bar a$.

To implement the recursion $\B$ stores a stack collecting the $b$'s of the
loops and the answer bits generated by the recursion as described. Scanning
the whole stack allows us to determine in space $O(\log n)$ the formula
$\psi(\bar x,\bar y)$ and tuple $\bar b$ such that the corresponding
recursive call is $\B(\psi(\bar x,\bar y),\str A,\bar a\bar b)$. The depth
of the recursion is at most $ \|\varphi\|$, so the stack can be stored in
space $\|\varphi\|\cdot (\log|A|+1)$. On an atomic formula $\B$ needs space
$\log |\str A|+\log |\varphi|\le O(\log n)$ for the loops on $j$ and $i$,
and again $O(\log n)$ for the equality checks. Altogether we see that~$\B$
can be implemented within the claimed space.
\end{proof}

The following proposition is the heart of the argument. The advantage with
respect to the ``brute force'' algorithm from the previous proposition is
that the factor $\|\varphi\|$ in the space bound is replaced by
$\log\|\varphi\|$. But since other factors are worsened this algorithm is
not in general more space efficient.

\begin{prop}\label{prop:sigma1}
There is an algorithm deciding $\textsc{mc}(\Sigma_1)$ that on an
instance $(\varphi,\str A)$ of~$\textsc{mc}(\Sigma_1)$ runs in space
\begin{equation}\label{eq:sigma1}
O\big( \log \|\varphi\|\cdot w(\varphi)\cdot\log |A|+ \log\|\varphi\|\cdot\log|\varphi|+ \log |\str A|\big).
\end{equation}
\end{prop}

\begin{proof}
We describe an algorithm $\A$ deciding the problem
\nnprob{a $\Sigma_1$-formula $\varphi$, a natural number $w\ge w(\varphi)$,
a structure $\str A$ and $\bar a\in A^{w}$}{$\str A\models \varphi(\bar{a})$?}
\noindent on an instance $(\varphi,w,\str A,\bar a)$ of length $n$ in
\emph{allowed space}
\begin{equation*}
O\big(\log \|\varphi\|\cdot w\cdot \log |A|+ \log\|\varphi\|\cdot\log|\varphi|+\log n\big).
\end{equation*}
Notationally, $\str A\models\varphi(\bar a)$ means that the assignment that
maps the $i$-th 
free variable in $\varphi$ to the $i$-th component of $\bar a$ satisfies
$\varphi$ in $\str A$. Here we suppose an order on the variables
in~$\varphi$, say according to appearance in $\varphi$. All we need is that
the value assigned to a given variable in a given subformula of $\varphi$
according to a given tuple $\bar a\in A^w$ can be determined in space~$O(\log
n)$.


For a sufficiently large constant $c\in\N$ to be determined in the course of
the proof, $\A$ checks that
\begin{equation}\label{eq:c}
\|\varphi\|\ge c\cdot w+c.
\end{equation}
If this is not the case, then $\A$ uses ``brute force'', that is,
it runs the algorithm from Lemma~\ref{lem:oed}.

Now suppose~\eqref{eq:c} holds. Choosing $c\ge 3$ this implies that the
syntax tree of $\varphi$ has at least 3 nodes. Then $\A$ computes in space
$O(\log |\varphi|)$ a subformula $\varphi_0$ of $\varphi$ such that
\begin{equation}\label{eq:spira}
\|\varphi\|/3\le \|\varphi_0\|\le 2\|\varphi\|/3.
\end{equation}
Observe that $\varphi_0$ is a $ \Sigma_1$-formula. Let $\bar y=y_1\cdots
y_{|\bar y|}$ list the free variables of $\varphi_0$ and note
\begin{equation}\label{eq:wy}
|\bar y|\le w(\varphi_0)\le w(\varphi)\le w.
\end{equation}

Recall from the intuitive sketch of the proof that we intend to call $\A$
recursively on a ``rest'' formula~$\varphi_1^{\bar b}$ where $\bar b$ is a
$|\bar y|$-tuple from $A$. To define this formula, let $c_1, \ldots,
c_{\max\{|\bar y|,1\}}$ be new constant symbols. For every free variable
$y_i$ of $\varphi_0$ check whether it has an occurrence in~$\varphi_0$ which
is not a free occurrence in $\varphi$. If such an occurrence exists, all the
free occurrences of $y_i$ in~$\varphi_0$ appear within a uniquely determined
subformula $\exists y_i \chi$ of $\varphi$ containing~$\varphi_0$ as a
subformula, where $\exists y_i$ binds these occurrences, that is, the free
occurrences of~$y_i$ in $\varphi_0$ are also free in $\chi$. Replace in
$\varphi$ the subformula $\exists y_i \chi$ by $\exists y_i(y_i{=}c_i \wedge
\chi)$. Let~$\varphi_1$ denote the resulting formula. Note that~$\varphi_1$
does not depend on the order of how these replacements for the $y_i$ are
performed.

Moreover, using~\eqref{eq:wy},
\begin{equation}\label{eq:phi1}
\big\|\varphi_1\big\|\le \|\varphi\|+ 2|\bar y|\le \|\varphi\|+ 2 w.
\end{equation}

\noindent
The algorithm $\mathbb A$ then loops through $\bar b=(b_1,\ldots,b_{|\bar y|})\in A^{|\bar y|}$
and does two recursive calls:
\begin{enumerate}[leftmargin=1cm]
\item[(R0)] Recursively call $\mathbb A(\varphi_0,w,\str A, \bar b)$ to
    check whether $\str A\models \varphi_0(\bar b)$. If $\str
    A\models\varphi_0(\bar b)$, then replace the subformula $\varphi_0$ in $\varphi_1$
    by $c_1{=}c_1$; otherwise by $\neg c_1{=} c_1$.

    Let $\varphi^{\bar b}_1$ be the resulting formula. Further, let $\str
    A^{\bar b}$ be the expansion of $\str A$ that interprets the constants
    $c_1,\ldots,c_{|\bar y|}$ by $b_1,\ldots, b_{|\bar y|}$ respectively.

\item[(R1)] Recursively call $\mathbb A\big(\varphi^{\bar b}_1,w,\str
    A^{\bar b}, \bar a\big)$ and output its answer.
\end{enumerate}

\noindent
Note that in (R1) we have $w(\varphi_1^{\bar b})\le w(\varphi)\le w$, so the
algorithm recurses to an instance of our problem. It is routine to verify
that
\begin{eqnarray*}
\str A\models \varphi(\bar a)
 & \iff &
 \text{there exists $\bar b\in A^{|\bar y|}$ such that $\str A^{\bar b}\models \varphi_1^{\bar b}(\bar a)$}.
\end{eqnarray*}

Thus, $\A$ correctly decides whether $\str A\models\varphi(\bar a)$. It
remains to show that $\A$ can be implemented in the allowed space.

We first estimate the depth of the recursion. In (R0) the algorithm recurses
to formula~$\varphi_0$ and $\|\varphi_0\|\le 2\|\varphi\|/3$ by~\eqref{eq:spira}. In (R1) the algorithm recurses on $\varphi_1^{\bar b}$, a
formula obtained from $\varphi_1$ by replacing the subformula $\varphi_0$ by
an atomic formula or the negation of an atomic formula. Then
\[
\|\varphi_1^{\bar b}\|\le \|\varphi\|+2w-\|\varphi_0\|+2\le
2\|\varphi\|/3+2w+2 = 3\|\varphi\|/4 + (2w+2- \|\varphi\|/12)
\]
where the inequalities hold by~\eqref{eq:phi1} and~\eqref{eq:spira},
respectively. Provided $c$ in~\eqref{eq:c} is large enough, this implies
$\|\varphi_1^{\bar b}\|\le3\|\varphi\|/4$. It follows that the recursion
depth is $O(\log\|\varphi\|)$.

In each recursive call the structure the algorithm recurses to is determined
by a tuple~$\bar b$ from the loop. This tuple has length at most $ w$
(cf.~\eqref{eq:wy}). The formula $\varphi_1^{\bar b}$ in (R1) is determined
by the truth value of $\str A\models\varphi_0(\bar b)$. The algorithm
recurses either
\begin{enumerate}[leftmargin=1cm]
\item[(P0)] to the formula $\varphi_0$, or

\item[(P1)] to the formula $\varphi_1^{\bar b}$ as defined if $\str
    A\models\varphi_0(\bar b)$, or

\item[(P2)] to the formula $\varphi_1^{\bar b}$ as defined if $\str
    A\not\models\varphi_0(\bar b)$.
 \end{enumerate}
To implement the recursion,~$\A$ maintains a sequence of tuples $\bar b_1,
\ldots, \bar b_d$ and a sequence of ``possibilities'' $(p_1,\ldots,
p_d)\in{\{0,1,2\}}^d$. The length $d$ is bounded by $O(\log\|\varphi\|)$, the
depth of the recursion. In a recursive call as described above the sequence
of tuples is expanded by the tuple~$\bar b$ from the loop, and the
possibility sequence by $0,1,2$ depending on which recursion from (P0),
(P1), (P2) is taken. Both sequences can be stored in allowed space, namely
\[
O(d\cdot w\cdot\log|A|)\le O(\log\|\varphi\|\cdot w\cdot\log|A|).
\]

We still have to explain how, given $(\bar b_1, \ldots, \bar b_d)$ and
$(p_1,\ldots, p_d)$, the algorithm determines the corresponding formula
$\psi$ and structure $\str B$ it has to recurse to. The structure $\str B$ is
an expansion of the input structure $\str A$ by a sequence of constants
interpreted by the sequence $\bar b_{i_1}\cdots\bar b_{i_\ell}$ where
$i_1<\cdots< i_\ell$ list those indices $j\in[d]$ such that $p_{j}\neq 0$.
This structure can be computed from the two sequences and the input in
logarithmic space. Note that $\str B$ is an expansion of $\str A$ by
interpreting at most $ w\cdot d$ many constants, so
\begin{equation}\label{eq:sizeB}
\log|\str B|\le \log(|\str A|+2\cdot w\cdot d)\le \log |\str A|+ \log w+\log \log \|\varphi\|+O(1).
\end{equation}

To determine the formula $\psi$, consider the function $R$ that maps a
formula $\chi$ and a number $i\le 2$ to the formula according to (P$i$). Since
we defined the recursion possibilities (P$i$) only on formulas
satisfying~\eqref{eq:c}, let us agree that $R(\chi,i):=\chi$ on formulas
violating~\eqref{eq:c}.
The desired formula $\psi$ is obtained by iterating this function along
$(p_1,\ldots, p_d)$: compute $\psi_1:=R(\varphi,p_1)$, $\psi_2:=
R(\psi_1,p_2)$, \ldots\ and output $\psi:=\psi_{d}$. Note that all these
formulas have length $O(|\varphi|)$, and each iteration step is computable in
space $O(\log |\varphi|)$. Hence, the whole iteration can be implemented in
space
\begin{equation*}
O(d\cdot \log|\varphi|)\le O(\log\|\varphi
\|\cdot\log|\varphi|).
\end{equation*}

If $(\psi,w,\str B,\bar b)$ is such that $\|\psi\|$ violates~\eqref{eq:c},
i.\,e.\ $\|\psi\|<c\cdot w+c$, then $\A$ invokes the ``brute force'' algorithm
from Lemma~\ref{lem:oed}. This requires space
\begin{equation*}\label{eq:leaf}
O\big(w\cdot\log|A|+\log |\psi|+\log|\str B|\big).
\end{equation*}
By $|\psi|\le O(|\varphi|)$ and~\eqref{eq:sizeB}, this is allowed space.
\end{proof}

\begin{rem}\label{rem:rev}
Recall Example~\ref{ex:fostconrel} defines a pl-reduction from
$p\textsc{-\STCon}$ to $\pmc(\Sigma_1^2)$ that maps an instance $(\str
G,s,t,k)$ of $p\textsc{-\STCon}$ to an instance $(\str A,\varphi)$ of
$\pmc(\Sigma_1^2)$ with $|\varphi|\le O(k)$. Combining with the algorithm
from Proposition~\ref{prop:sigma1} we thus decide $p\textsc{-\STCon}$ in
space
\begin{equation}\label{eq:ksav}
O(\log k\cdot\log |G|).
\end{equation}
This is a slightly more detailed statement of Savitch's
Theorem~\ref{thm:savitch}. As pointed out by an anonymous reviewer, a kind of
converse holds, namely, Proposition~\ref{prop:sigma1} can be derived from
Savitch's Theorem by means of a reduction:
\end{rem}

\begin{proof}[Sketch of a second proof of Proposition~\ref{prop:sigma1}.]
Given an instance $(\str A,\varphi)$ of $\pmc(\Sigma_1)$ construct the
following directed graph $\str G$. Its vertices~$G$ are triples
$(\psi,\alpha,b)$ where $b\in\{0,1\}$ is a bit and~$\psi,\alpha$ are as in
the proof of Theorem~\ref{thm:mccomplete}: $\psi$ is a subformula of
$\varphi$ and $\alpha$ is an assignment of its free variables. Again we
assume that negations appear only in front of atoms. Note
\begin{equation}\label{eq:revgraph}
|G|\le 2\cdot \|\varphi\|\cdot |A|^{w(\varphi)}.
\end{equation}

The goal is to define the edges of $\str G$ in such a way that there is a
path from $s:=(\varphi,\emptyset,0)$ to $t:=(\varphi,\emptyset,1)$ in $\str
G$ if and only if $\str A\models\varphi$. Moreover, if this is the case,
then there is such a path of length at most $k:=\|\varphi\|$. This allows to
decide whether $\str A\models\varphi$ by running Savitch's algorithm on
$(\str G,s,t,k)$. By~\eqref{eq:ksav} and~\eqref{eq:revgraph} this needs
space~\eqref{eq:sigma1} and thus proves Proposition~\ref{prop:sigma1} (it
will be clear that this space suffices to construct $\str G$).

It remains to define the edges of $\str G$. If $\psi$ is an atom or a
negated atom, add an edge from $(\psi,\alpha,0)$ to $(\psi,\alpha,1)$ if and
only if $\alpha$ satisfies $\psi$ in $\str A$.

If $\psi$ is $(\psi_0\vee\psi_1)$, add edges from $(\psi,\alpha,0)$ to
$(\psi_0,\alpha_0,0)$ and $(\psi_1,\alpha_1,0)$, and add edges from
$(\psi_0,\alpha_0,1)$ and $(\psi_1,\alpha_1,1)$ to $(\psi,\alpha,1)$; here
$\alpha_0$ and $\alpha_1$ are the restrictions of $\alpha$ to the free
variables of $\psi_0$ and $\psi_1$, respectively.

If $\psi$ is $(\psi_0\wedge\psi_1)$, add edges from $(\psi,\alpha,0)$ to
$(\psi_0,\alpha_0,0)$, from $(\psi_0,\alpha_0,1)$ to $(\psi_1,\alpha_1,0)$,
and from $(\psi_1,\alpha_1,0)$ to $(\psi,\alpha,1)$; here $\alpha_0,\alpha_1$
are defined as in the previous case.

If $\psi$ is $\exists y\chi$, then add, for every $a\in A$, edges from
$(\psi,\alpha,0)$ to $(\chi,\beta_a, 0)$ and from $(\chi,\beta_a, 1)$ to
$(\psi,\alpha,1)$; here, $\beta_a$ equals $\alpha$ if $y$ is not free in
$\chi$, and otherwise extends $\alpha$ by mapping $y$ to~$a$.
\end{proof}

We now extend the space bound from the previous proposition to
$\textsc{mc}(\Sigma_t)$ for each $t\ge 1$.

\begin{proof}[Proof of Theorem~\ref{thm:sigmat}.]
Similarly as in the previous proposition, we give an algorithm $\A$ deciding
\nnprob{a $\Sigma_t$-formula $\varphi$, a natural number $w\ge w(\varphi)$, a
structure $\str A$ and $\bar a\in A^{w}$} {$\str A\models \varphi(\bar a)$?}
\noindent on an instance $(\varphi,w,\str A,\bar a)$ of length $n$ in
\emph{allowed space}
\begin{equation*}
O\big(\log \|\varphi\|\cdot w\cdot \log |A|+ \log\|\varphi\|\cdot\log|\varphi|+\log n\big).
\end{equation*}
For $t=1$ this is what has been shown in the proof of
Proposition~\ref{prop:sigma1}. So we assume $t\ge 2$ and proceed
inductively.

Let $\psi_1,\ldots,\psi_r\in\Pi_{t-1}$ be such that $\varphi$ results from
these formulas by existential quantification and positive Boolean
combinations. Of course, such formulas $\psi_j$ are computable in
logarithmic space from $\varphi$. For $j\in[r]$ let $s_j\le w(\varphi)\le w$
be the number of variables occurring freely in $\psi_j$ and let $\bar x_j$
be a length $s_j$ tuple listing these variables.

Let $\str A^*$ be the structure with universe $A$ that interprets for each
$j\in[r]$ an $s_j$-ary relation symbol $R_j$ by
\[
R^{\str A^*}_j:=\{\bar b\in A^{s_j}\mid \str A\models\psi_j(\bar b)\}.
\]

By the induction hypothesis we can compute $\str A^*$ in allowed space.
Moreover
\begin{equation}\label{eq:sizeAstar}
|\str A^*|\le \|\varphi\|+ |A|+ \|\varphi\|\cdot |A|^{w}.
\end{equation}

Define the formula $\varphi^*$ by replacing for every $j\in[r]$ the formula
$\psi_j(\bar x_j)$ in $\varphi$ by $R_{j}(\bar x_j)$. Clearly, we have
$\varphi^*\in \Sigma_1$
\begin{equation}\label{eq:sizevarphistar}
\|\varphi^*\|\le \|\varphi\|\quad \textup{ and }\quad |\varphi^*|\le O( |\varphi|).
\end{equation}
More importantly,
\begin{eqnarray*}
\str A\models \varphi(\bar a)
 & \iff &
\str A^*\models \varphi^*(\bar a).
\end{eqnarray*}

Since $\varphi^*$ is a $\Sigma_1$-formula with $w(\varphi^*)\le
w(\varphi)\le w$ we have that $(\varphi^*,w,\str A^*,\bar a)$ is an instance
of the problem treated in the proof of Proposition~\ref{prop:sigma1}. We can
thus decide whether $\str A^*\models\varphi^*(\bar a)$ in space
\[
O\big(\log \|\varphi^*\|\cdot w\cdot \log |A|+ \log\|\varphi^*\|\cdot\log|\varphi^*|+\log 
|\str A^*|\big)
\]
By~\eqref{eq:sizevarphistar} and~\eqref{eq:sizeAstar}, this is allowed space.
\end{proof}

\section{Summary and future directions}

We have studied the parameterized space complexity of model-checking bounded
variable first-order logic, i.\,e.\ $\pmc(\FO^s)$ for fixed $s\ge 2$. We
stratified the problem into subproblems according quantifier alternation rank
and showed (Theorem~\ref{thm:mccomplete}) that the respective subproblems
$\pmc(\Sigma^s_t)$ are complete for the levels of the tree-hierarchy
$\TREE[t]$, the alternation hierarchy above the class \TREE\ from~\cite{cm1}.
We further showed that allowing function symbols does not increase the space
complexity of these problems (Theorem~\ref{thm:func}). This gives a quite
fine-grained picture of the space complexities
 up to pl-reductions.

However, it is open whether the tree-hierarchy is strict even under some
plausible complexity assumption. We showed it does not collapse to $\paraL$
if Savitch's Theorem is optimal, in fact, then already $\PATH\neq\paraL$
(Theorem~\ref{thm:improvesavitch}). It follows that $\pmc(\Sigma^2_1)$
cannot be solved in parameterized logarithmic space if Savitch's Theorem is
optimal. Under this assumption we proved a stronger result
(Theorem~\ref{thm:upper}) stating, intuitively, that the na\"{\i}ve
model-checking algorithm is space-optimal.

Finally, Theorem~\ref{thm:upper} gives a highly space-efficient
model-checking algorithm for $\pmc(\Sigma^s_t)$. We presented two
constructions, a direct one and another, pointed out to us by an anonymous
reviewer, via a reduction of $\pmc(\Sigma_1^2)$ to $p$-$\stdipath_\le$ and
Savitch's algorithm.

\medskip
We view the results about the tree hierarchy as a contribution to the
fine-structure theory of~FPT~(cf.~\cite{m}). In fact,
Theorems~\ref{theo:class} and~\ref{thm:completeness} indicate that the tree
hierarchy contains many natural parameterized problems. However, as pointed
out by an anonymous reviewer, Theorem~\ref{thm:completeness}~(2) implies
that problems in $\TREE[*]$ have shallow circuits, and hence, intuitively,
$\TREE[*]$ is a small subclass of FPT\@. Parameterized circuit complexity is
another emerging theory about the fine-structure of FPT~\cite{est,bst,bantan,cftree,mrdp,cfac}.

We repeat the questions whether $\PATH$ or $\TREE$ are closed under
complementation. We noted that a positive answer for $\TREE$ would imply a
collapse of the tree hierarchy (Corollary~\ref{cor:collapse}). We do not know
whether this also follows from $\PATH$ being closed under complementation.

As a further structural question we do not know how $\paraNL$ relates to the
tree hierarchy. Proposition~\ref{prop:nlvstreestar}~(1) gives only a partial
answer.

We showed that the straightforward model-checking algorithm is space-optimal
under the hypothesis that Savitch's Theorem is optimal
(Theorem~\ref{thm:lower}). We conjecture that similar optimality results can
be derived for other natural algorithms as well. Future work will show to
what extent the hypothesis that Savitch's Theorem is optimal can play a role
in space complexity similar to the one played by the ETH in time complexity.

\section*{Acknowledgements}
We thank the anonymous reviewers for their detailed comments and critical
suggestions on how to better present our work. In particular, we are grateful
for pointing out the elegant reduction from the second proof of
Proposition~\ref{prop:sigma1}.

\bibliographystyle{alpha}
\bibliography{CEM}

\end{document}